\documentclass[12pt]{article}

\usepackage{a4}
\usepackage{epsf}
\usepackage[dcu]{harvard}

\citationmode{abbr}
\citationstyle{dcu}

\newcommand{\eq}{\begin{equation}}
\newcommand{\en}{\end{equation}}
\newcommand{\eqa}{\begin{eqnarray}}
\newcommand{\ena}{\end{eqnarray}}

\begin{document}

\hskip 8.0truecm In press, {\it Neural Computation}.

\bigskip \bigskip \bigskip

\centerline{\bf \Large
A unified approach to the study of }
\vskip 0.1cm 
\centerline{\bf \Large temporal, correlational and
rate coding } 

\bigskip

\centerline{\large Stefano Panzeri$^{\P}$ and Simon R. Schultz$^{\S}$}
\vskip 0.1cm
\bigskip
\centerline{\it \large $^{\P}$ Neural Systems Group, University of Newcastle upon Tyne }
\vskip 0.1cm
\centerline{\it \large Department of Psychology, Ridley Building}
\vskip 0.1cm
\centerline{\it \large $^{\S}$ Howard Hughes Medical Institute and }
\vskip 0.1cm
\centerline{\it \large Center for Neural Science, New York University }
\vskip 0.1cm
\centerline{\it \large 4 Washington Place, New York, NY 10003, U.S.A. }


\newpage

\begin{abstract}
We demonstrate that the information contained in the spike occurrence
times of a population of neurons can be broken up into a series of
terms, each of which reflect something about potential coding
mechanisms. This is possible in the coding r{\'e}gime in which few
spikes are emitted in the relevant time window. 

This approach allows us to study the additional information
contributed by spike timing beyond that present in the spike counts;
to examine the contributions to the whole information of different
statistical properties of spike trains, such as firing rates and
correlation functions; and forms the basis for a new quantitative
procedure for the analysis of simultaneous multiple neuron
recordings. It also provides theoretical constraints upon neural
coding strategies. We find a transition between two coding
r{\'e}gimes, depending upon the size of the relevant observation
timescale. For time windows shorter than the timescale of the
stimulus-induced response fluctuations, there exists a spike count
coding phase, where the purely temporal information is of third order
in time. For time windows much longer than the characteristic
timescale, there can be additional timing information of first order,
leading to a temporal coding phase in which timing information may
affect the instantaneous information rate.

In this new framework we study the relative contributions of the
dynamic firing rate and correlation variables to the full temporal
information; the interaction of signal and noise correlations in
temporal coding; synergy between spikes and between cells; and the
effect of refractoriness. We illustrate the utility of the technique
by analysis of a few cells from the rat barrel cortex.
\end{abstract}

\newpage

\section{Introduction}

Information about sensory stimulation is, at the most fundamental
level, represented in the central nervous system by the spike emission
times of populations of neurons. In principle, the temporal pattern of
spikes across the neuronal population provides a large capacity for
fast information transmission \cite{Mac+52}. It is still unclear how
much of this theoretical capacity is actually exploited by the brain.

It has long been known that a substantial amount of sensory
information is carried by the discharge rate of individual neurons
\cite{Adr26}. In some circumstances, however, if
the stimulus is modulated on a very short time scale, precisely
replicable sequences of spikes can be obtained
\cite{Bai+96,Bur+98}. Does this represent temporal coding, or is the
relevant time window for counting spikes merely very short? Recent
analyses have also suggested that temporally coded information is
present in the spike trains of individual neurons in the monkey visual
cortex under more general stimulation conditions
\cite{Vic+96tcod,Vic+98,Mec+98}. Evidence has also accrued that some information
appears to be encoded by stimulus (or behaviour) related changes in
the coordination of timing of firing between small populations of
cortical cells (\citename{Vaa+95}, 1995; \citename{deC+96}, 1996;
\citename{Riehle+97}, 1997). Given that our understanding of how spike 
trains are decoded biophysically is far from complete, the rigorous
study of the information properties of nerve cells requires that we
can quantify spike train information in full generality. Understanding
neural coding then means understanding how the different features of
the population of spike trains (such as spike counts, correlations or
patterns) contribute to the full temporal information. This provides a
powerful constraint upon the biophysical decoding that occurs: for
instance, if no significant information about a stimulus is present in
feature X of the spike trains, then none can be decoded by recipient
neuronal pools by using `X detection'.

To study these questions, it is of interest to quantify information
using a rigorous measure such as the mutual information
\cite{Sha48,Cov+91}. In this context, the Shannon mutual information
measures the extent to which observing a spike train (or a number of
spike trains from several cells) reduces the uncertainty as to which
external stimulus was present -- it provides a bound upon well the
stimuli can be discriminated. Equivalently, it measures the fidelity
of coding on a trial by trial basis -- how reproducibly the responses
on individual trials represent the stimulus. If the spike times are
observed with finite temporal precision $\Delta t$, and if $\Delta t$
is small enough such that all timing fluctuations below that timescale
are not influenced by stimulation, then a binary string can be formed
(with binwidth $\Delta t$) which contains all of the information
present in the spike train. Direct estimation of the full temporal
information is in principle possible by measuring the frequency of
occurrence of all possible patterns of ones and zeros (`words') that
the string can take from the experimental data
\cite{Str+98,deR+97}. A direct approach is particularly useful because
it does not make any assumptions about what are the important aspects
of the spike train, and is thus completely general. However, direct
estimation by `brute-force' estimation of frequencies is problematic,
because of the large data sizes required (growing up to exponentially
with wordlength). Despite this limitation, a recent study has
succeeded in directly quantifying full temporal information from an
alert animal, in part because of the low spiking rates obtained under
the stimulation conditions used \cite{Bur+98}.

Another approach to direct calculation of the information is to expand
the mutual information as a Taylor series in the experimental time window
\cite{Bia+91,Ska+93,Pan+96speed,Pan+99cor}. This is the
approach taken here. In this paper we extend previous work based on
the spike count response for a small population of cells
\cite{Pan+99cor} to present for the first time an analytical
expression for the full temporal information in a population of spike
trains to second order in the time window. This enables a natural
separation of the contributions of both instantaneous firing rates and
temporal correlations between spikes to the complete information. It
also allows comparison with the information carried just by the number
of spikes fired by each cell, which can be expressed in an equivalent
form. This provides theoretical insight into the individual factors
which determine the coding capabilities of neural spike trains. It
also provides a procedure for neurophysiological data analysis which is superior with respect to data size requirements
compared to `brute force' frequency estimation of the information.

The use of such a Taylor series approach requires that the
experimental time window (which should not be confused with the
smaller binwidth used to discretise the spike train) be short enough
that only a few spikes are emitted in its duration. This short time
window limit, although a restriction, is relevant to the transmission
of sensory information in the mammalian cortex. Single unit recording
studies in primates have demonstrated that the majority of information
about static visual stimuli is often transmitted in windows as short
as 20-50ms \cite{Tov+93,Hel+95,Rol+99mask}.  Information about
time-dependent signals is often conveyed by single sensory cells by
producing about one spike per characteristic time of stimulus
variation \cite{Rie+96}. Event related potential studies of the human
visual system \cite{Tho+96} provide further evidence that the
processing of information in a multiple stage neural system can be
extremely rapid. The periods during which transcranial magnetic
stimulation disrupts processing in early visual cortex have been found
to be as short as 40 ms \cite{Cor+99}. Finally, the assessment of the
information content of interacting assemblies which may last for only
a few tens of milliseconds \cite{Sin+97} requires the use of such
short windows.

The paper is organised as follows: in section 2, the problem will be
defined, and the series expansion of the information explained. In
section 3, the necessary rate and correlation parameters will be
introduced, and the probabilities of each possible response expressed
in terms of these parameters. Section 4 gives the main result of the
paper, the analytical expression obtained by substituting these
probabilities back into the equation for mutual information. In
section 5, we examine the conditions under which this approach is
valid. Section 6 uses these results to study the role of response
timescales in neural coding; two distinct coding r{\'e}gimes are
found, depending upon the relative stimulus and response
characteristic timescales. Temporal encoding may contribute
significantly when the mean instantaneous firing rates of the cells
fluctuate on a timescale shorter than the time window in which most of
the information about the stimulus is transmitted. Section 7 examines
the precision with which spike times may code information. In section
8, we study the conditions under which one finds synergistic
relationships between spikes and between cells in an assembly. This
includes an analysis of the effect of a refractory period on the spike
train information. In section 9, we illustrate the method by applying
it to spike trains recorded in the rat barrel cortex. Finally, in
section 10, we discuss the consequences of the results and their
relationship to other work.

\section{The information carried by neural spike trains}

Consider a time period of duration $T$, associated with a dynamic or
static sensory stimulus, during which we observe the activity of $C$
cells. This period of physiological observation has associated with it
a period (which might be earlier by some `lag' time, but which we will
consider to have the same length) in which we also observe the
characteristics of an external correlate, such as a sensory stimulus,
which might be influencing the cells' behaviour.  Let us denote each
different stimulus history (time course of characteristics within $T$)
as $s(\tau)$, a function of time from stimulus onset chosen from the
set ${\cal S}$ of experimentally presented stimulus histories. We
shall describe the neuronal population response to the stimulus by the
collection of spike arrival times $\{t^a_i\}$, each $t^a_i$ being the
time of occurrence of the $i$-th spike emitted by the $a$-th neuron.

Although the spike arrival time is a continuous variable, it can be
experimentally measured only with finite precision $\Delta t$. Let us
divide the time window $T$ into small bins of width $\Delta t$ (in
which at most one spike per cell is observed). With complete
generality, we can represent the spike sequence $\{t^a_i\}$ as a
sequence of binary digits, one for each time bin and cell, with the
response ``1'' in the bins corresponding to the spike times, and ``0''
in all other bins. We denote the probability of observing a spike
sequence $\{t^a_i\}$ when a particular stimulus history $s(\tau)$ was
present as $P(\{t^a_i\}|s(\tau))$
\footnote{\normalsize We use $P(\cdot)$ to indicate a probability, and
$p(\cdot)$ a probability density.}. $P(\{t^a_i\}) =
\left<P(\{t^a_i\}|s(\tau))\right>_s$ is its average across all
stimulus histories. We determine $P(\{t^a_i\}|s(\tau))$ by repeating
each stimulus history in exactly the same way on many trials, and
observing the responses.

Following \citeasnoun{Sha48}, we can write down the mutual information
provided by the spike trains about the whole set of stimuli as
\eqa 
I(\{t_i^a\}; {\cal S}) &=& \int {\cal D}s ~p[s(\tau)]
\int {\cal D}t_i^a 
p\left[\{t_i^a\}|s(\tau)\right]\log_2 {p\left[\{t_i^a\}|s(\tau)\right]
\over p(\{t_i^a\})}
\nonumber\\
&\equiv& \sum_{s(\tau) \in \cal S} P[s(\tau)]
\sum_{t_i^a}
P\left[\{t_i^a\}|s(\tau)\right]\log_2 {P\left[\{t_i^a\}|s(\tau)\right]
\over P(\{t_i^a\})} .
\label{eq:tempmutinfo} 
\ena
This notation can be read as follows. In the first, more general form
of the equation, we use the functional integral notation $\int {\cal
D}s$ to indicate that in principal, a continuous set of stimulus
histories can be created. In practice, a discrete and often very
limited set of stimuli is usually used to test the cell -- hence we
replace the functional integral with a summation over a discrete set
of stimuli. For brevity, the $\tau$ will henceforth be dropped unless
there is a particular need to stress the dynamic nature of the
stimulus function. The notation $\int {\cal D}t_i^a$ indicates
integration over all spike times $t_i^a, i=1..n_a, a=1..C$, and
summation over all total spike counts from the population. $n_a$ is
the number of spikes emitted by cell $a$. This notation is very
similar to that utilised in \cite[p. 158]{Rie+96}. In the limit of
infinite precision, $\int {\cal D}t_i^a$ may be interpreted as a
simple Riemannian integral; the results we present in this paper are
valid in this continuous limit. More usually we will invoke
finite precision $\Delta t$ and interpret the integral as a summation
over all time bins.

In addition to studying the information contained in the sequence of
spike times, we can also quantify the information contained in the
response space defined by only the number of spikes emitted by each
cell in the time window. We can form a $C$-dimensional vector ${\bf n}$, each
component of which is $n_a$, the number of spikes fired by the respective
neuron in the experimental time window. The spike count information
can be written
\eq
I({\bf n}; {\cal S}) =  \sum_{s \in \cal S} P(s) \sum_{{\bf n}} P({\bf
n}|s)\log_2 {P({\bf n}|s) \over P({\bf n})} .
\label{eq:countmutinfo}
\en 
This information quantity is exactly that calculated in
\cite{Pan+99cor}.
Invoking the data processing inequality, it is relatively obvious that
$I({\bf n}; {\cal S}) \le I(\{t_i^a\}; {\cal S})$.

Now, the spike train information can be approximated by a power series
\eq I(\{t_i^a\}; {\cal S}) =  I_t(\{t_i^a\}; {\cal S}) \; T +
I_{tt}(\{t_i^a\}; {\cal S}) \; {T^2\over 2}\;  + \dots 
\label{eq:taylorinfo}
\en where $I_t(\cdot)$ and $I_{tt}(\cdot)$ refer to the first and
second time derivatives of the information respectively. As we shall
see later, this becomes effectively an expansion in a dimensionless
quantity, the number of spikes fired in the time window. For this
approximation to be valid, the information function must be analytic
in $T$, and the series must converge within a few terms. Both of
these issues will be addressed in section~5. The time derivatives of
the information can be calculated by taking advantage of the
narrowness of the time windows, as will be explained.

\section{Correlation functions and response probability}

Before specifying the expressions for the response probabilities, it
is necessary to define some response parameters. Consider first the
discrete time resolution case (finite $\Delta t$). For each
individual trial with stimulus $s$, the spike train density of each
cell can be represented as a sum of pulses,
\eq
r_a(t;s) = \sum_i {\delta_{t,t_i^a} \over \Delta t} ,
\label{spikedensdiscr}
\en
where $ \delta_{t_1,t_2}$ is the Kroenecker delta function (1 if $t_1$
and $t_2$ label the same time bin, and zero otherwise). $i$ in the
above indexes the spike number. The time-dependent firing rate is
measured as the average of this quantity over all experimental trials
with the same stimulus $s$. We denote this trial average by a bar, and
the firing rate is thus $\overline{r}_{a}(t ; s)$. Note that this mean
rate function is sometimes called the post-stimulus time histogram, or
PSTH.

It is also necessary to introduce parameters describing correlations
among spikes. (Note that we use the word ``correlation'' to include
both autocorrelation and crosscorrelation). Now, it is apparent that,
for physically plausible processes, $\overline{r}$ will remain
constant as $\Delta t \to 0$: if the mean firing rate is a
differentiable function of time, then for sufficiently short bin
widths, reducing $\Delta t$ will just reduce the probability of
observing a spike accordingly. We would like to retain this property
for our correlation measure as well, so that when we write out the
series expansion of the information equation, there will not be any
dependence upon time hidden inside the response parameters. Such
hidden dependence would prevent a simple intuition of the dependence
of the relative magnitudes of the information terms upon the
timescale, and would, the authors, feel, be inelegant. The Pearson
product moment, for instance, can be shown to approach zero for short
time windows
\cite{Pan+99cor}. The measure we choose is the scaled correlation
density. We can write the scaled noise correlation density, which
measures correlations in the response variability upon repeated trials
of the same stimulus\footnote{\normalsize We adopt the convention, utilised by a
number of authors, of using the term `noise correlation' to mean
correlation {\em at fixed signal}, thus distinguishing it from
correlation across different signals, which concept we will also
require.}{ as
\eqa 
\gamma_{ab}( t^a_i , t^b_j  ; s)  &=& \frac{\overline{r_{a}(t_i^a ; s)
r_{b}(t_j^b; s)}}{\overline{r}_{a}(t_i^a;
s)\overline{r}_{b}(t_j^b ; s)} -1 , \; {\rm if} ~ a \neq b ~ {\rm
or} ~ t_i^a \neq t_j^b \nonumber \\ 
 \gamma_{aa}( t^a_i , t^a_i  ; s)  &=& -1 .
\label{gammameasure}
\ena
The numerator of the first term indicates the average, over trials in
which the same stimulus $s$ is presented, of the product of the spike
densities of cell $a$ at time $t_i^a$ and cell $b$ at time $t_j^b$.
Note that the scaled correlation density is simply a joint
post-stimulus time histogram (JPSTH) from which the number of
coincidences purely due to rate modulation has been subtracted
(implemented by the ``-1'' in equation~\ref{gammameasure}).

Now, we can introduce the above correlation parameters in terms of the
conditional firing probabilities of observing one spike from cell $a$
in the time bin centered at $t^a_i$, given that cell $b$ emitted a
spike in the time bin centered at $t^b_j$, when stimulus $s$ was
presented:
 \begin{equation}
P(t^a_i | t^b_j ; s) \equiv \overline{r}_{a}(t^a_i ; s) \;
\Delta t [ 1 + \gamma_{ab}( t^a_i , t^b_j  ; s) ] + O(\Delta t^2).
\label{conditionalprobability}
\end{equation}
This assumes that the conditional probabilities
(\ref{conditionalprobability}) scale proportionally to $\Delta t$. It
is a natural assumption, as it merely implies that the probability of
observing a spike in a time bin is proportional to the resolution
$\Delta t$ of the measurement. It is violated only in the implausible
and non-physical case of spikes locked to one another with infinite
time precision. It can of course be checked for any given dataset;
this will be carried out in section~5. Note that quadratic (and
higher) order terms in $\Delta t$ were neglected from the above
relationship as they affect only third and higher order terms of the
information. 

We will find it convenient to measure another type of correlation
also: signal correlation, that is correlation in the mean responses of the
neurons across the set of stimuli. These correlations can also be
thought of as correlations in the tuning curves of the neurons. For
homogeneity, we will also quantify these by a scaled correlation density:
\begin{equation}
\nu_{ab}(t^a_i , t^b_j) = 
{ \left< \overline{r}_a(t^a_i; s) \overline{r}_b(t^b_j ; s)\right>_s \over
\left<\overline {r}_a(t^a_i; s)\right>_{s} \left<\overline {r}_b(t^b_j
; s)\right>_{s} } -1 . 
\label{signalcorr}
\end{equation}
In the above, the brackets $\left<\cdot\right>_s$ can be taken to
indicate the average across stimuli, $\int {\cal D}s~ p(s) \cdot$ or
$\sum_s P(s) \cdot$. Both $\gamma$ and $\nu$ may range from -1 to
$\infty$, with zero indicating lack of correlation.

When passing to the high resolution limit ($\Delta t \to 0$), the same 
definitions outlined above apply, provided that we replace the
Kronecker delta function with that of Dirac in the definition of spike 
density:
\eq
r_a(t;s) = \sum_i \delta(t - t_i^a) ,
\label{spikedenscont}
\en
and use probability densities instead of probabilities in
equation~\ref{conditionalprobability}.

Let us now consider the case of the spike count response
parameters. The correlational parameters that influence the spike
count information are the scaled correlation coefficients of the spike
counts in each trial \cite{Pan+99cor}. These are obtained by summation
over time bins (or integration in the high resolution limit) of the
above expressions. The spike count scaled noise correlation
coefficient can be written
\eq
\gamma_{ab}(s) = {\int dt_i^a \int dt_j^b ~
\overline{r}_{a}(t_i^a ; s) \overline{r}_{b}(t_j^b; s) [1 +
\gamma_{ab}(t^a_i, t^b_j , s)] \over [\int dt_i ^a
~\overline{r}_{a}(t_i^a; s)][ \int dt_j^b ~\overline{r}_{b}(t_j^b ; s)]}
-1 .
\label{countgammameasure}
\en

Similarly, signal correlation is quantified as:
\begin{equation}
\nu_{ab} = { \left< \int dt_i^a ~\overline{r}_a(t^a_i; s) \int dt_j^b
~\overline{r}_b(t^b_j ; s)\right>_s \over \left< \int dt_i^a ~\overline{r}_a(t^a_i;
s)\right>_{s} \left<\int dt_j^b ~\overline{r}_b(t^b_j ; s)\right>_{s}
} -1 .
\label{countsignalcorr}
\end{equation}
Equations~\ref{countgammameasure},\ref{countsignalcorr} are a simple
renotation of the definitions in \cite{Pan+99cor} to fit with the
requirements of the full temporal notation in the current paper. The
finite resolution expressions for the above are of course obtained by
replacing the integrals with summations: \eq \lim_{\Delta t
\rightarrow 0}
\sum_n f(t_n)
\Delta t = \int dt f(t)
\label{eq:discrtocont}
\en

If the conditional instantaneous firing rates are non-divergent, as
assumed in equation \ref{conditionalprobability}, then the short time
scale expansion of response probabilities becomes essentially an
expansion in the total number of spikes emitted by the population in
response to a stimulus. The only responses which contribute to the
transmitted information up to order $k$ are the responses with up to
$k$ spikes from the population. This is proved in Appendix A. The
only relevant events for the second order analysis are therefore those
with no more than two spikes emitted in total, and they can be
truncated at second order without affecting the first two information
derivatives. This is valid for any time-resolution. For brevity, we
report the results only for the infinite time resolution case. The
resulting expression (see Appendix A) is
\begin{eqnarray}
P({\bf 0}|s) & = & 1 - \; \sum_{a=1}^C \int dt_1^a
\overline{r}_{a}(t_1^a ; s) + \nonumber \\  
&+& {1 \over 2}  \sum_{a=1}^C  \sum_{b=1}^C \int dt_1^a \int dt_2^b 
\overline{r}_{a}(t_1^a ; s) \overline{r}_{b}(t_2^b ; s) \left[1 + 
\gamma_{ab}(t_1^a , t_2^b ; s)\right]    
\label{prob_0} \nonumber\\
p(t^a_1|s) dt^a_1 &= &  \overline{r}_{a}(t_1^a ; s) dt^a_1 \left(1 -  
\sum_{b=1}^C \int dt_2^b
\overline{r}_{b}(t_2^b ; s)
\left[1 + \gamma_{ab}(t_1^a , t_2^b ; s)\right] \right) \phantom{ppp} a =
1,\cdots,C \label{prob_1} \nonumber\\ 
p(t_1^a t_2^b |s) dt_1^a dt_2^b &= & {1\over 2} \overline{r}_{a}(t_1^a
; s) \overline{r}_{b}(t_2^b ; s) \left[1 +  
\gamma_{ab}(t_1^a , t_2^b ; s)\right] dt_1^a dt_2^b
\phantom{ppp} a,b = 1,\cdots,C ; 
\label{eq:resp_prob}
\end{eqnarray}
where $P({\bf 0}|s)$ is the probability of zero response (no cells
fire), $p(t^a_1|s)$ is the probability density of observing just one
spike from cell $a$ at the specified time location, and $ p(t_1^a
t_2^b |s) $ is the probability density of observing just a pair of spikes
at the given times.

\section{Analytical results} 

We now insert the second order response probabilities
(equation~\ref{eq:resp_prob}) into the expressions for the information
contained in the sequence of spike times (equation~\ref{eq:tempmutinfo})
and the spike counts (equation~\ref{eq:countmutinfo}). Then, for each
term in the sum over responses, we utilise the power expansion of the
logarithm as a function of $T$,
\eq
\log_2(1 - T \; x) = - {1\over \ln 2} \sum_{j=1}^\infty {(T
x)^j \over j} ,
\label{eq:logexpansion}
\en
and, after integrating over spike times, group together all terms in
the sum which have the same power of $T$. Equating these with
equation~\ref{eq:taylorinfo} yields expressions for the information
derivatives. These expressions depend only upon the time-dependent
firing rates $\overline r$, the noise correlations $\gamma$, and
the signal correlations $\nu$. As shown in Appendix A, the power expansion in the short time window length $T$ is related to the expansion in the total number of spikes emitted by the population. Therefore, although the formalism is for simplicity developed as an expansion in $T$, the expansion parameter is in fact the adimensional quantity representing the average number of spikes emitted by the population in the window $T$.

The first order contribution (i.e. $T$ times the instantaneous
information rate) to the full temporal information from a
population of spike trains is
\cite{Bia+91}
\eq I_t(\{t_i^a\}; {\cal S}) \; T =
\sum_{a=1}^C \int dt^a \left< \overline{r}_{a}(t^a ; s) \log_2
{\overline{r}_{a}(t^a ; s)\over \left<\overline{r}_{a}(t^a ;
s')\right>_{s'}}~\right>_{s} .
\label{eq:temp1stderiv}
\en
This is simply a sum of single cell contributions. It is insensitive
to both signal and noise correlation.

The expression for the second order contribution (i.e. the second temporal
derivative multiplied by $T^2/2$) breaks up into three terms:
\eqa
I_{tt}(\{t_i^a\}; {\cal S}) \; {T^2 \over 2} & = & {1\over 2\ln 2}
\sum_{a=1}^C \sum_{b=1}^C \int dt_1^a \int dt_2^b \left<\overline
{r}_{a}(t_1^a ; s)\right>_{s} \left<\overline {r}_{b}(t_2^b ;
s)\right>_{s} \nonumber \\ & &\times \biggl\{ \nu_{ab}(t_1^a , t_2^b) +
\left[1 + \nu_{ab}(t_1^a , t_2^b)\right]\ln {1\over 1+\nu_{ab}(t_1^a,t_2^b)}
\biggr\} \nonumber \\ & + & {1\over 2} \sum_{a=1}^C \sum_{b=1}^C \int
dt_1^a \int dt_2^b \left< \overline{r}_{a}(t_1^a ; s)
\overline{r}_{b}(t_2^b ; s) \gamma_{ab}(t_1^a , t_2^b ; s) \right>_s
 \log_2 {1\over 1+\nu_{ab}(t_1^a , t_2^b)} \nonumber \\ & + &
{1\over 2} \sum_{a=1}^C \sum_{b=1}^C \int dt_1^a \int dt_2^b \biggl< \overline{r}_{a}(t_1^a ; s) \overline{r}_{b}(t_2^b ; s) \left[1 +
\gamma_{ab}(t_1^a , t_2^b ; s)\right] \nonumber \\ & &\times \log_2
\biggl\{{
\left<\overline{r}_{a}(t_1^a ; s') \overline{r}_{b}(t_2^b ;
s')\right>_{s'}\left[1+ \gamma_{ab}(t_1^a , t_2^b ; s)\right]\over \left<\overline{r}_{a}(t_1^a ; s')
\overline{r}_{b}(t_2^b ; s') \left[1+ \gamma_{ab}(t_1^a , t_2^b ;
s')\right]\right>_{s'} } \biggr\}\biggr>_s .
\label{eq:temp2ndderiv}
\ena
We will refer to these terms as components {\em 2a, 2b}\/ and {\em 2c}
of the information respectively.

When each spike is completely independent, as in a Poisson process, it
is apparent that only the first term of equation~\ref{eq:temp2ndderiv}
survives. It is easy to see that this term is always less than or
equal to zero, since $f(x) = x-(1+x)\ln(1+x)$ has a global maximum at
$f(0) = 0$: the first term is equal to zero only if the signal
correlation is precisely zero. This means that the information
accumulation from a Poisson process with $|\nu|>0$ always slows down
after the first spike.  

If there is any deviation from independence in the timing of
successive spikes, then the other terms can contribute to the
information through non-zero autocorrelation density
$\gamma_{aa}(t_1^a , t_2^b)$. If there is any relationship between the
times of spike emission of different cells, then they can contribute
through non-zero cross-correlation $\gamma_{ab}(t_1^a , t_2^b)$. As
with the spike count information terms detailed in \cite{Pan+99cor},
the second of these terms (the ``stimulus-independent correlational
component'') reflects contributions from a level of correlation which
is not stimulus dependent. The third term is non-negative, and is non-zero only in the
presence of stimulus dependence of the correlation between spikes, and
is called the ``stimulus-dependent correlational component''. The
natural separation of the second order information into three
components is important because each component reflects the
contribution of a different relevant encoding mechanism.  When
applying this analysis to real neuronal data, a significant amount of
information found in the third component of $I_{tt}$, relative to the
total information, would clearly signal that cells are transmitting
information mainly by participating in a stimulus (or context)
dependent correlational assembly \cite{Sin+97}. A specific example
will be given in Figure~\ref{fig:fastcor}.

In the same way, the short time scale expansion can be carried out for
the spike count information. This was performed in detail in
\citeasnoun{Pan+99cor}, and here we report briefly the main results to
cast them into identical notation to that above for  
comparison.  The spike count information derivatives are similar to
equation~\ref{eq:temp2ndderiv}, but they depend only on the mean rate across
time and on the spike count correlation coefficients
(\ref{countgammameasure},\ref{countsignalcorr}). The first order
contribution is:
\eq 
I_t({\bf n}; {\cal S}) \; T =  \sum_{a=1}^C 
\left< \int dt^a \overline{r}_{a}(t^a ; s) \log_2 {\int dt^a \overline{r}_{a}(t^a ; s)\over
\left< \int dt^a \overline{r}_{a}(t^a ; s')\right>_{s'}}~\right>_{s}
\label{eq:count1stderiv}
\en 

As before, the second order contribution is broken into three components:
\eqa
I_{tt}({\bf n}; {\cal S}) \; {T^2 \over  2} & = & {1\over 2\ln 2} \sum_{a=1}^C  \sum_{b=1}^C 
\left<\int dt_1^a \overline {r}_{a}(t_1^a ; s)\right>_{s} \left<\int dt_2^b 
\overline {r}_{b}(t_2^b ; s)\right>_{s}  \nonumber 
\\
& \times& \biggl[  
\nu_{ab} + (1 + \nu_{ab})\ln {1\over 1+\nu_{ab}}  \biggr]
\nonumber \\
& + &  {1\over 2} \sum_{a=1}^C  \sum_{b=1}^C \left< \left( \int dt_1^a \overline{r}_{a}(t_1^a ; s) \right)
\left( \int dt_2^b \overline{r}_{b}(t_2^b ; s)  \right)
\gamma_{ab}(s) \right>_s 
 \log_2 {1\over 1+\nu_{ab}} \nonumber \\
& + & {1\over 2} \sum_{a=1}^C  \sum_{b=1}^C 
\biggl< \left(\int dt_1^a \overline{r}_{a}(t_1^a ; s)\right) 
\left( \int dt_2^b \overline{r}_{b}(t_2^b ; s) \right) \left[1 + 
\gamma_{ab}(s)\right]
\nonumber \\ 
& \times &
\log_2 \biggl\{{ \left< \int dt_1^a \overline{r}_{a}(t_1^a ; s')
\int dt_2^b \overline{r}_{b}(t_2^b ; s')\right>_{s'}\left[1+ \gamma_{ab}(s)\right]\over  
\left< \int dt_1^a \overline{r}_{a}(t_1^a ; s')
\int dt_2^b \overline{r}_{b}(t_2^b ; s') \left[1+ \gamma_{ab}(s')\right]\right>_{s'} } \biggr\} \biggr>_s.
\label{eq:count2ndderiv}
\ena

\section{Limitations}
\label{sec:roval}

In this section we address the potential limitations of the technique
presented here as a procedure for the analysis of real data. There are 
four assumptions which must be satisfied for the series approximation
to be guaranteed to be a good estimate of the true information. These
are: that the experimental time window is small; that the conditional
firing probability scales with $\Delta t$; that the information is an
analytic function of time; and that experimental trials are
statistically indistinguishable.

{\bf Assumption 1}. The short time scale limit utilized here formally
requires that the mean number of spikes in the time window be
small. The actual range of validity of the order $T^2$
approximation will depend on how well the time dependence of the
information from the neuronal population fits a quadratic
approximation. The range of validity for the spike count series
information was checked by simulation in \cite{Pan+99cor}. The same
range of validity must hold for the temporal information as well when
the firing rates and correlations of the neurons vary slowly with
respect to the time window considered, since in this limit the
additional temporal information is small (see next section). The
additional numerical test required here is therefore on the range of
applicability of the second order approximation for the full spike
timing information when the parameters describing the neuronal firing
probabilities fluctuate quite rapidly within the time window $T$.

For this purpose we simulated neurons driven by two stimuli. The
response of each neuron to each stimulus was modeled as a 1 ms
resolution Poisson process with a time dependent firing rate (the
response in each 1 ms time bin is generated with a time dependent
firing rate independently of the response in the other time bins). The
first stimulus gave a constant (flat) mean response $r_0 = 30$
spikes/sec. across time, whereas the second stimulus was chosen to
produce a response of mean $r_0$ and sinusoidally modulated with
amplitude $r_0$, at a frequency $1/\tau_c$. After a full period, the
two stimuli are indistinguishable on the basis of spike count alone,
and therefore this is a good model for examining temporal coding.

The analytical approximations to the information were tested against
the true information computed directly from the model response
probabilities. Figure ~\ref{fig:roval1cell} shows the accuracy of the
first and second order approximations to the information in the spike
times for a single cell with responses to the second stimulus
oscillating very fast (at a frequency of 500 Hz). The second order
approximation is nearly exact up to 200 ms, even in the extreme case
of such fast fluctuations. The range of validity seems not to be
significantly affected by changing the frequency of oscillations. We
have also carried out simulations with a set of two cells with either
Poisson or cross-correlated firing. When simulating two cells, we
could not systematically test time windows as long as 200 ms because
the number of spike trains over which we have to sum to compute the
full information increases exponentially both with the length of the
window and with the population size. However, we found that the
scaling expectation that the range of validity shrinks as $1/C$ with
population size is confirmed up to the time window lengths that we
could test. 

The simulations presented here therefore show that the series
expansion can useful for time scales relevant to neuronal coding, for
the small ensembles of cells typically recorded during {\em in vivo}
experimental sessions.  In particular, when correlations between
spikes are weak the series expansion is very precise up to hundreds of
ms. However, one should be aware that the presence of very strong
correlations between the spikes can potentially reduce this range of
validity. Therefore the actual range of validity for the data under
analysis should be further assessed by looking e.g. at the agreement
between the series expansion result for spike counts and the brute
force evaluation of the spike count information from response
probabilities (equation~\ref{eq:countmutinfo}). (The latter, unlike
the brute force full temporal information, can usually be computed
with an experimentally feasible number of trials). An example of this
check is given in section 9.

{\bf Assumption 2}. The second assumption required is that the
probability of observing a spike in the time bin centered upon $t_i^a$
from cell $a$ given that one has been observed in the time bin centred
upon $t_j^b$ from cell b scales with $\Delta t$. This is captured by
equation~\ref{conditionalprobability}. This assumption would be
expected to break down only if there were a significant number of spikes
synchronised with near-infinite precision.

To illustrate how this scaling assumption can be validated experimentally, we examined the scaling relation between the average value of $P(t^a_i
| t^b_j ; s)$ and $\Delta t$ for two neurons recorded from the rat barrel cortex. The stimuli were upward deflections of one of 9 whiskers. This data was
provided by M.E. Diamond and colleagues. For further reference to the data see 
\cite{Leb+00} and section 9, were the information properties of the same two cells are analyzed with our method. The result can be seen in
Figure~\ref{fig:scaling}. We found that the averaged conditional firing rate, computed as the average across all bin pairs and stimuli of $P(t^a_i
| t^b_j ; s)/\Delta t$, remained approximately constant as $\Delta t$ was decreased, and no divergence of the conditional firing rate was observed for all the resolutions considered. This provides evidence that Assumption 2 holds for this dataset.

{\bf Assumption 3}. 
This method expands the mutual information as a Taylor series in $T$. Therefore the third required assumption is that the information is an analytic function of the time window size $T$. 
This requirement ensures that the Taylor series expansion of the mutual information exists, and that the series converges to the mutual information.

The validity of this assumption can be tested, at least to some extent, on the dataset under analysis. For example, one can check that there are no discontinuities in time of the firing parameters. Also, if there are enough data, the mutual information can be computed by brute force from the Shannon formula, equation~\ref{eq:tempmutinfo}, although it can be usually computed only for a more limited time range than when using the series expansion (see section 9). Hence the information computed by brute force from the Shannon formula can be compared, in some time windows, to the series expansion estimation. This provides a strong check of the validity of this particular assumption and of the convergence of the series expansion method, as fully illustrated in section 9. It is apparent that the assumption is valid for the data we examined here (see section 9). This assumption is more likely to cause problems with artificially constructed models implementing instantaneous correlations between neurons. One could argue that such models are non-physical in any case.

{\bf Assumption 4}. The final assumption required is that experimental
trials are statistically indistinguishable. This assumption is of
course inherent to any neurophysiological data analysis, and is taken
to be the definition of an experimental trial. Satisfaction of this
assumption is the domain of experimental design.

\section{Response timescales and pure temporal coding} 

Having quantified the information $I(\{t_i^a\}; {\cal S})$ in
the spike times and $I({\bf n}; {\cal S})$ in the spike counts as a
function of the rates and correlations, we are in a position to study
the conditions under which information is not dominated by the spike
counts. In other words, we ask: how much extra information, not
contained in the spike counts, is conveyed by the temporal relations
between spikes. This extra information is precisely $I(\{t_i^a\};
{\cal S}) - I({\bf n}; {\cal S})$.

We find that the crucial parameter is the typical timescale of
variation of the firing rate and correlation functions, relative to
the time window $T$ of interest. We will label this typical or
characteristic stimulus-induced response timescale $\tau_c$.

{\bf Case 1, large $\tau_c$}. If the scale of time variation of any
time dependent function $f(t)$ is large compared to the time window
considered, then $f(t)$ can be approximated by its power expansion and
the following quasi-static approximation holds:
\eq
\int_{t_0 - T/2}^{t_0 + T/2} f(t) dt = f(t_0) \;T + {\partial^2 \over
\partial t^2} f(t)|_{t=t_0} \; {T^3 \over 24} + O(T^5) 
\label{slowapprox}
\en As a consequence, if both rate and correlation functions vary
slowly in the time domain of interest, only the rate and correlation
values near the centre of the interval ($t_0$) are important, and the
extra amount of information in the spike times is sub-leading (only of
order $T^3$). The expression for the $T^3$ term in the extra
information in spike timing can be computed by applying
equation~\ref{slowapprox}) to the integrals over time involved in
computing information rates (equations~\ref{eq:temp1stderiv} and
\ref{eq:count1stderiv}). The result depends only on the firing rate
variations near $t_0$, and not on correlations between the spikes.
\eqa
& & I(\{t_i^a\}; {\cal S}) - I({\bf n}; {\cal S}) = \nonumber \\ 
& & {T^3 \over 24 \ln 2}  \sum_{s} P(s) {\partial  \over \partial t} r(t ; s)|_{t = t_0}  \nonumber \\ 
& & \left( 
{(\partial  / \partial t) r(t ; s)|_{t = t_0} \over r(t_0 ; s)} - {<(\partial  / \partial t) r(t ; s')|_{t = t_0}>_{s'} \over < r(t_0 ; s')>_{s'}}
\right) .
\label{extraslow}
\ena 
(\ref{extraslow}) can be shown to be non-negative for any rate
function, as required by information theoretical consistency.

Being of order $T^3$, in this r{\'e}gime pure temporal coding affects
neither the rate nor acceleration of information transmission. Pure
temporal information would accumulate very slowly with time. Therefore
under these circumstances information is dominated by the spike
counts.

{\bf Case 2, small $\tau_c$}. If either firing rates or correlation
functions fluctuate with a characteristic timescale smaller than $T$,
then the quasi-static approximation (\ref{slowapprox}) breaks down,
and there is a transition to a different coding r{\'e}gime. 
A possible case may be the presence of a sinusoidal component of period shorter than $T$ in the rate fluctuations. 
If in particular $\tau_c$ is {\em much} shorter than $T$, the integral of the function over the time window does not
depend on the value of the function at the center of the interval as
in equation \ref{slowapprox}, or on the frequency of response variations,
but only on the average value of the function over the period $\tau_c$:
\eq
\int_{t_0 - T/2}^{t_0+T/2} f(t) dt = \left[ {1\over\tau_c}
\int_{\tau_c} f(t) dt \right] \;T \; + O(T^2) 
\label{fastapprox}
\en

As a consequence, in this limit the additional purely
temporal information is proportional to $T$ (and not to $T^3$ as in
the quasistatic phase):
\eqa
I(\{t_i^a\}; {\cal S}) - I({\bf n}; {\cal S}) &=&
\sum_{a=1}^C \int dt^a 
\left< \overline{r}_{a}(t^a ; s) \log_2 {\overline{r}_{a}(t^a ; s)\over
\left<\overline{r}_{a}(t^a ; s')\right>_{s'}}~\right>_{s}
\nonumber \\
&-&
\sum_{a=1}^C 
\left< \int dt^a \overline{r}_{a}(t^a ; s) \log_2 {\int dt^a \overline{r}_{a}(t^a ; s)\over
\left< \int dt^a \overline{r}_{a}(t^a ; s')\right>_{s'}}~\right>_{s}
\nonumber\\ 
&+& O(T^2)
\ena
This means that when rates fluctuate rapidly the actual {\em rate} of
information transmission (measured in bits/sec) can be considerably
faster than that with spike counts only. Thus in this phase there can
be substantial pure temporal information, and it may even be dominant
with respect to the spike count information.

We studied particular cases by simulating mean firing rate and
correlation parameters for each timestep as described in
section~\ref{sec:roval}. Two stimuli were again used, once inducing a
flat rate response function, and the second a sinusoidally modulated
function of time. The following results were obtained with 1 ms time
resolution, but have also been confirmed by numerical integration in
the infinite time-resolution limit. The effect of timing precision is
considered separately. Figure~\ref{fig:cf} illustrates a situation in
which the spike count of a single cell responding according to a
Poisson process is unable to discriminate between the stimuli after a
full period, but such information is available from spike timing. The
time window is increased, beginning at the onset of
stimulus-related responses. In this case, after 60 ms (a full cycle of
oscillation), the spike count information drops to zero, whereas the
full temporal information accumulates cycle after cycle. After one
cycle the information is purely temporal.

The effect of stimulus modulation frequency on the full temporal
information is shown in Figure~\ref{fig:freq}a. Figure~\ref{fig:freq}b
shows the purely temporal information for the same situation (i.e. the
full information minus the spike count information). The difference
between the fast and slow r{\'e}gimes can be seen by comparing
these two figures.  Slow fluctuations lead to negligible (of order
$T^3$) pure temporal information. In the fast regime, pure temporal
information increases roughly proportionally to the window
length. Note also that, if fluctuations are very fast ($\tau_c \ll T$)
and the spike incidence is measured with high temporal precision, the
amount of information is roughly independent of the frequency of
oscillation, as predicted by (\ref{fastapprox}).

If the rates vary slowly with respect to $T$, but the correlations
vary on a faster time scale, then the pure timing information can be
only of second order in $T$. This means that in this situation the
instantaneous rate of information transmission is unaffected by the
temporal structure of the spike train. However, the temporal
information can still be appreciable. An example is shown in
Figure~\ref{fig:fastcor}. This Figure plots the information in the
responses of a pair of cells, each of which has the same
characteristics as those in Figures~\ref{fig:cf} and
\ref{fig:freq}. Both of the cells' firing rates are slowly (1 Hz)
modulated by one stimulus only. However, the correlation between them
is modulated more rapidly, with a frequency of 100 Hz, and it is
stimulus dependent.  The correlation for one stimulus is equal in
magnitude but opposite in sign to that of the other stimulus: the
correlation between the cells is modulated at $f=100$ Hz according to
$\gamma_{ab}(t^a , t^b ; s) = \pm \exp (- | t^a - t^b|/\lambda)
\sin(2\pi f t^a) \sin(2 \pi f t^b)$. The decay constant
$\lambda$ was chosen to be 25 ms. This particular cross-correlation
function was chosen to match the one used in Figure~\ref{fig:popsyn}, but the result
is indicative of the behaviour of any fast-oscillating
cross-correlation. There is, as can be seen, a substantial
contribution to information component $2c$, the third, stimulus dependent,
component of the second order information. This contribution has no
counterpart in the spike count information, which remains much
smaller.

\section{Precision of spike timing}

The total information contained in the spike times, equation~\ref{eq:tempmutinfo}, is also a function of the precision $\Delta t$
with which the spikes are measured (or equivalently, of the precision
in the spike timing itself). Experimental measures of the information
with different values of $\Delta t$ can address the question of to
what temporal precision information is transmitted in the cerebral
cortex. The whole question of whether spike timing is important is
really the question of whether the use of time resolution as short as
a few milliseconds significantly increases the information extracted
\cite{Str+98}.  Measuring information at high resolutions with a brute
force direct evaluation from equation~\ref{eq:tempmutinfo} is made difficult
by the exponential increase of the number of trials needed as $\Delta
t$ is decreased (see \citeasnoun{Str+98} and Appendix B). Our
formalism gives a simple expression for the information in spike
timing as a function of the timing precision $\Delta t$ (the discrete
version of equations~\ref{eq:temp1stderiv} and \ref{eq:temp2ndderiv}),
and requires only a quadratically increasing number of trials as
$\Delta t \rightarrow 0$.  Thus it can be used to measure the
information with resolutions that would be impractical with brute
force methods using data sets of the size of typical cortical
recording sessions \cite{Pan+96}.

As an example, Figure~\ref{fig:precis} illustrates the impact of
sampling precision on the full temporal information contained in 32 ms from a
single model cell responding to two stimuli according to a Poisson process. 
The first stimulus elicits a constant response firing rate, the second stimulus elicits a response firing rate oscillating sinusoidally in time, exactly as in Figure 3. 
In Figure~\ref{fig:precis}, the oscillation frequencies for the responses to the second stimulus were varied in the range 5-250 Hz. 
The result is that the timing precision has no effect on the information
only if it is much smaller than the typical time scale of response
parameter variations $\tau_c$. If $\Delta t \simeq \tau_c$, then
information is strongly underestimated with respect to the infinite
resolution limit.

The issue of the efficiency of information transmission for
different timing precisions (i.e. how much of the total entropy is
actually exploited for information transmission) is addressed in
\citeasnoun{Sch+00tempent}.

\section{Synergy and redundancy in temporal information}

How is information combined from a group of coding elements?  Is the
amount of information obtained from the whole pool of elements greater
than the sum of that from each individual element (synergistic) or
less than the sum (redundant)? Between these two cases, there can
exist a situation where the information from each element is
independent, and the total information thus increases linearly as the
number of elements is increased.  Two notions of ``synergy'' are of
immediate relevance. The first is synergy/redundancy between
assemblies of cells; the second is synergy between spikes (whether
they are from the same cell or another).

{\bf Synergy between cells}. We define the amount of synergy between
cells as the total information from the ensemble of spike trains minus
the sum of that from the individual cells \cite[p. 268]{Rie+96}. The
redundancy is simply the negative of this quantity. This means that,
to second order, the synergy is simply the sum of the off-diagonal ($a
\neq b$) elements of equation~\ref{eq:temp2ndderiv}. This is because the
first order terms (equation~\ref{eq:temp1stderiv}) of the information
and the $a=b$ terms in the summation are identical for both the ensemble
information and the sum of the individual cell informations, thus
eliminating in the subtraction.

The amount of synergy between cells depends critically upon
correlation. For synergistic coding, non-zero cross-correlation is
needed. This can be seen from equation~\ref{eq:temp2ndderiv}: in
the absence of correlation $\gamma$  only the first term of $I_{tt}$ would
survive, and we have shown this term to be less than or equal to
zero, meaning that synergy cannot occur.

It is quite obvious that synergy may occur when the degree of
correlation between cells is modulated by the stimulus. However, even when 
noise correlation is not stimulus dependent, it is possible
to achieve synergistic coding. This is done by using the second,
stimulus-independent, correlational component of $I_{tt}$ to increase
the total information. This happens, for each pair of cells, when the
noise correlation $ \gamma_{ab}( t^a_i , t^b_j ; s)$ is opposite
in sign to the signal correlation $\nu_{ab}( t^a_i , t^b_j )$ for
times $ t^a_i , t^b_j$. This basic mechanism of synergy for the
temporal information is identical in principle to that considered for
spikes counts in \cite{Ora+98,Abb+99,Pan+99cor}; it extends naturally
to noise and signal correlations between time pairs $t^a_i$ and $t^b_j$.

Correlations between cortical neurons often oscillate and vary rapidly
in sign with time \cite{Kon+95pnas}. Our analysis shows that in this
case, to obtain a maximally synergistic effect, the signal correlation
should oscillate in counterphase with respect to the cross-correlogram
and with similar frequency. This produces opposite signs of signal and
noise correlation. If the signal was of lower frequency, the effect of
cross-correlation would be washed away by its rapid change of sign
with respect to the signal. The effect of the relative frequencies and
phase of signal and noise variations with time is illustrated in
Figure~\ref{fig:popsyn}. We simulated two Poisson cells, responding to
two stimuli as in Figure~\ref{fig:cf}a, but with firing rate in response to the
second stimulus being modulated at $f_1 = 10$ and $f_2 = 20$ Hz respectively. The
signal correlation between the cells is in this simple case
\eq
\nu(t^a,t^b) \propto \sin(2\pi f_1 t^a) \sin(2 \pi f_2 t^b).
\en
To generate a cross-correlation model that could be tuned to
match the signal correlation frequency, the cross-correlation was chosen
for both stimuli to be of the form 
\eq
\gamma_{ab}(t^a , t^b ; s) = \gamma
exp (- | t^a - t^b|/\lambda) \sin(2\pi k f_1 t^a) \sin(2 \pi k f_2
t^b) .
\en
The cross-correlation function decayed exponentially, with time
constant 25 ms, as a function of the time between spikes
\cite{Kon+95pnas}. The constant $\gamma$ in
front of the correlation function was set either to 0 (no
cross-correlation), to -1 (signal and noise in counter-phase), or to 1
(signal in phase with the noise). $k$ determined the relative
timescales of signal and noise correlation oscillation, and was set to
1 for Figure~\ref{fig:popsyn} part A and 10 for part B. The
information from the pair of cells observed simultaneously is compared
with the summed single cell information in the Figure.

When the noise correlation between the cells is chosen to oscillate
with the same frequency as the signal correlation does, synergy is
obtained when signal and noise correlation are in
counter-phase. Conversely high redundancy is obtained when they are
in phase. When instead, signal and noise oscillate on very different
time scales, correlations play a much less significant role. 
When response profiles of different neurons are independent,
and little signal correlation is present, weakly stimulus modulated
correlations do not affect information transmission at all
\cite{Ora+98,Pan+99cor}. This is important as signal correlation
between cortical neurons was reported to be very small for stimulus
sets of increasing complexity \cite{Gaw+96,Rol+97ebr,Dea+99},
and therefore correlations might be less important under natural
conditions than when using artificial and limited laboratory stimuli.

The neuronal model used in this simulation is certainly far from
realistic; however, the result discussed here is valid beyond the
simple models used for the Figures.  We note that when
cross-correlations are stimulus dependent, then the third term of
equation~\ref{eq:temp2ndderiv} also becomes non-zero, and additional
interactions between response parameters can arise. They can be
studied by specifying the stimulus dependence of correlations and
taking into account also this information component.
 
In conclusion, this section shows that the contribution of cross-correlations to the representation of the external world can be more complex than what
might be expected from naive visual inspection of
cross-correlograms. This reinforces the need for a rigorous
information theoretic analysis of the role of trial-by-trial
correlations in solving complex encoding problems, like feature
binding \cite{Sin+97}.

\subsection*{Synergy between spikes}

Another notion of synergy, defined for single cells (as well as
populations) and for the full temporal information, is synergy between
{\em spikes}. Motivated by the notion that particular sequences of
spikes may have a special role in encoding stimuli,
\citeasnoun{Bre+99} recently introduced a synergy measure of this
type.  In that work, ``events'' (e.g. single spikes, or pairs of
spikes occurring with a given time delay irrespective of whether there
are other spikes in-between) are singled out from the whole spike
train; the information carried by occurrence of single events was
computed\footnote{\normalsize The definition of
information carried by the occurrence of a single event is a
generalization of equation~\ref{eq:temp1stderiv}. However, the
\citeasnoun{Bre+99} definition of information carried by, for example, single
spikes in isolation is different from the contribution of the
occurrence of a single spike in a given trial to the information
from the full spike train. The latter quantity also has terms
of higher order.}. \citename{Bre+99}
considered the synergy to be the information from a complex event
(e.g. a pair of spikes occurring at specified times) minus the
information from each of those individual spikes constituting that
pair.

Another way to measure the synergy between spikes that is suggested by
the present analysis is to take into account their contribution to the
information conveyed by the whole spike train. The contribution of
single spikes to the whole spike train information is given by
equation~\ref{eq:temp1stderiv}, and correspondingly the contribution
of pairs of spikes is given by the sum of
equations~\ref{eq:temp1stderiv} and \ref{eq:temp2ndderiv}. Therefore
the extent of synergy between pairs of spikes is simply the quantity
given in equation~\ref{eq:temp2ndderiv} alone. The concept generalises 
to higher order interactions; `spike' synergy is something which falls 
naturally out of the series expansion approach.

It is interesting to note that the introduction of a refractory
period has an accompanying effect on the synergy between spikes, as
measured by the fractional second order contribution to the overall
temporal information. Figure~\ref{fig:spksyn}a shows the effect of the
refractory period on the synergy between spikes. This is shown for
the example of a cell whose firing rate is modulated by a 40 Hz
sinusoid for one stimulus as in the earlier Figures. The Figure
compares a Poisson process with processes augmented with an absolute
refractory period of between 1 and 16 ms, and also with another
process which has a 2ms absolute refractory period, and an
exponentially decaying autocorrelation with time constant 20 ms,
reflecting a relative refractory period. It is apparent that
for the Poisson process (in which spikes are not entirely independent
because of the rate modulation) the second order contribution is small
or negative. For longer refractory periods there is an increasing peak
in the early period of the response. This effect appears to involve an
interaction between the absolute refractory period and the
time-constant of stimulus modulation of the rates; the stimulus
modulation timescale determines the width of the spike-synergy peak
(see Figure~\ref{fig:spksyn}b), whereas the refractory period length
determines its height. 

It has been noted previously that the temporal
correlations introduced into the spike train by refractoriness can
have a beneficial effect on coding. This occurs by regularization of
the higher firing rate parts of the response, producing a more
deterministic relation between stimulus and response
\cite{deR+97,Ber+98}. The present analysis confirms this effect, and
adds several new facts: that the effect is second order in time, and
that it involves interaction between the refractory and dominant
stimulus-induced modulation timescales, such that the width of the
period of increased information is determined by the stimulus
frequency characteristics, whereas the amount of extra information is
determined by the effective duration of refraction.

\section{Analysis of neurophysiological data}

In order to illustrate the specific advantages and the type of
neurophysiological results that can be obtained with our series
expansion method, we present here an example analysis of two cells
recorded from the barrel cortex of adult normal anaesthetised Wistar
rats. The two neurons presented were located in the D2 and C2 barrel
column respectively. Each of the stimulation consisted of a 100 ms
lasting upward deflection of one of the whiskers. For each neuron, the
stimulus set was composed of its principal whisker and its eight
surrounding whiskers, so that each of the whiskers that were likely to
activate the cells were included in the stimulus set. 50 to 56 trials
per whisker were available.  The time onset of stimulation, as
well as spike times, were recorded with 0.1 ms precision. (See
\citeasnoun{Leb+00} for a complete description of the experimental
methods). The information about which of the nine whiskers was
stimulated was computed, both with the series expansion approach and
using brute force estimation from response frequencies. Information
from spike counts and spike times were both computed. For the latter,
a resolution $\Delta t$ = 10 ms was used. Corrections for the upward
bias originating from limited sampling were applied (see Appendix
B). The range over which unbiased information measures could be
obtained is discussed below and in Appendix B. Note that the
probability scaling assumption was shown to be satisfied by both
neurons in Figure~\ref{fig:scaling}.

Figure~\ref{fig:exam1}a shows the spike count information analysis for
the first neuron. Since the overall mean rate of the neuron across the
100 ms stimulation is 10 Hz, the series expansion is expected to be
very good. Indeed, a comparison of the brute force and series spike
count information shows that the two are essentially identical for the
whole stimulation time range. As the two quantities are (after finite
sampling corrections) unbiased in this time range, this is a very
compelling verification of the validity of the series expansion
method. In Figure~\ref{fig:exam1}b we report the spike times information
analysis for the first neuron. It is evident that the brute force
estimation of the full temporal information diverges rapidly after the
first four to five time bins (i.e. after to 40-50 ms). This is due to
failure of corrections for finite sampling, as expected by the rule of
thumb for sampling corrections (see Appendix B). The spike times
series expansion is a close match to the brute force estimator up to 40-50 ms, and no appreciable divergence due to finite
sampling is visible, again as expected. This illustrates the clear
superiority, in terms of sampling requirements, of the series
expansion with respect to brute force evaluations. The separation of
information into components, obtained using the series expansion,
shows that most of the information is carried by the rate
component. However, for this cell spike correlations contribute up to
approximately 15\% of the information in the spike times case. Note
that this information in spike correlations is all contributed by the
stimulus-independent component. This means that this little
correlational information does not arise from modulation of the
autocorrelogram with stimulus, but from the interplay of signal and
noise correlations. The peak across time for the spike count
information was 0.31 bits; the full temporal information obtained with
10 ms resolution was of 0.41 bits at time T = 50 ms. This shows that a
neuron can transmit appreciable temporal information within a fraction
of one mean interspike interval. This possibility was predicted by our
mathematical analysis of coding regimes. Finally, it is interesting to
note that this neuron in the first 10 ms of response carried on
average more than 3 bits per each spike emitted.

Figure~\ref{fig:exam2} reports the information time course for the
second neuron. Its mean rate across 100 ms was 16 Hz. Also in this
second case the information for the spike counts obtained with the
series expansion coincided with that obtained by the brute force
method, thus confirming the reliability of the series expansion. When
considering spike times, series expansion and brute force evaluation
are very close up to 40-50 ms. After 50 ms, the brute force evaluation cannot be corrected effectively for finite sampling, and thus it starts to diverge. 
The peak
spike count information was 0.13 bits; the peak spike timing
information was 0.20 bits. As before, significant temporal information
is transmitted within one typical interspike interval. The component
analysis indicates that, unlike the first cell, this second cell transmits information by mean firing rate modulations only, and correlations between spikes do not transmit any information.

The above two cells were shown only for illustrative purposes, rather
than to make any general point about the information properties of
barrel cortex neurons\footnote{\normalsize A complete study of the informational properties of a large set of barrel cortical neurons will be the subject of a separate publication (Panzeri, Petersen, Schultz, Lebedev and Diamond; in preparation)}. 
This illustration
shows that i) the series expansion method can give reliable and
testable results, ii) when used to compute information in spike times,
it has considerable advantages in terms of data size requirements; and
iii) it can effectively quantify the contributions of different
encoding mechanisms to neuronal information transmission.

\section{Discussion}

We have demonstrated that the \cite{Mac+52} information contained in
the specification of the times of spike emission of a population of
cells can be broken up into a series of terms each quantifying the
contribution of a different encoding mechanism, just as the
information contained in the spike counts from a population can
\cite{Pan+99cor}. Although it can be applied only to a limited range
of time windows, the use of this series approach has a number of
advantages over brute force computation for both the understanding of
the theory of coding with spike trains, and for the analysis of
neurophysiological data.

A first advantage of this approach is that it necessarily compares the
contributions of different information-bearing parameters on correct
and equal terms, and shows how they combine to yield the full
information available from the spike train. A second advantage of the
series expansion method is that it requires much less data sampling
than a brute force approach, in order to obtain unbiased and reliable
information measures.  This is because the information is computed by
using only the subset of the possible variables characterising the
spike train which carry the most information in the short time
limit. Therefore the complexity of the response space is reduced in
the way that preserves the most information. A third interesting
feature of the method present here is that it does not assume that the
neuronal response space is a vector space, as e.g. Principal Component
Analysis \cite{Opt+87} does.  Since sensory systems are non-linear,
this is a property that has to be satisfied by any method for studying
neuronal information encoding \cite{Vic+96tcod,Vic+97Net}. Fourth, the
series expansion approach is a method to quantify the full temporal
information that can be extracted from the spike train by an ideal
observer, and does not depend on the validity of a stimulus-response
model, and is not specific to a particular stimulus decoder. Therefore
it provides information values against which the performance of
encoding/decoding models can be assessed \cite{Rie+96,Bor+99}.

The observation timescales with which spike trains are studied depend
upon the questions which are being asked. Many authors, particularly
those with a perspective from the field of psychophysics, have
examined windows of data up to several seconds long. The justification
is that, in certain tasks \cite{Bri+92}, psychophysical performance
increases up to these timescales, as does the ability to discriminate
stimuli based upon counting the spikes of a single cell (as counting
for longer allows noise to be averaged out). The techniques discussed
in the current paper are inapplicable to such long time windows
(although other techniques such as direct calculation of the spike
count information for a single cell do apply). The authors would like
to comment that the study of neural coding is often considered as a
precursor to, or partner of, the study of computation or information
processing in the nervous system. To understand the computational
architecture of the brain, it is necessary to understand the nature of
the symbols which are processed. Accepting that single neurons are the
primary information processing units of the brain, these symbols must
exist on timescales which are ``seen'' by single cells, which may be
as low as a membrane time-constant, or longer if additional
integrative processes are enacted. It is neural coding on single cell
timescales which determines computation, and which motivates the
current work. Since perceptual processes are unlikely to be uniform in
time, such a strategy might also elucidate some of the computations
underlying perception at longer timescales.

The work presented here is not only relevant as a tool for
neurophysiological data analysis. It allows also analytical studies of
the statistical properties of population spike trains, such as
refractoriness, autocorrelation, and timing relationships between
cells, to be conducted with regard to understanding which parts of the
full temporal information they contribute to and which they do not.
Here we have used this formalism to relate precisely the time scale of
response parameter variations to transitions between spike count and
temporal encoding regimes; to show that the impact on information
representations of stimulus-independent trial-by-trial correlations
depends crucially on their interplay with the signal correlation; and
to show that the neuronal refractory period can lead to synergy
between spikes.  A further understanding of how other parameters of
biophysical relevance can be adjusted to maximise encoding capacity
can be reached by the use of the formalism presented here.

One most interesting finding often reported in experimental studies of
neuronal information transmission is that sensory neurons transmit
most of the information within one mean interspike interval, and
single spikes carry a lot of information \cite{Rie+96}. Thus high
spike timing precision and rate coding may coexist in sensory neuron,
as the rate code has to be evaluated in a short window. This has led
some authors to conclude that under these conditions the distinction
between rate and temporal encoding may become blurred (see
\citeasnoun{Rie+96}, p. 119). However, differences between temporal
and rate encoding and rate coding can be defined and be present even
under conditions of such rapid processing \cite{Theu+95,Bor+99}. Our
study advances the understanding of the distinction between temporal
and rate encoding when the relevant window for information
transmission is short. In fact, our work explicitly relates response
variations time scales to transitions between coding regimes, and
shows that under some conditions a substantial amount of information
can be encoded purely temporally within a fraction of one interspike
interval (see section 6). Indeed, the example analysis of two neurons
in the rat barrel cortex shows that this is not only a theoretical
possibility, but that it can be realised by sensory neurons.
    
\citeasnoun{Sha+98} argue that because the interspike interval in the
responses of cortical neurons is highly variable, the rapid
information transmission achieved by the cerebral cortex (i.e.
substantial information being transmitted in one ISI or less) must
imply redundancy of signal.  Their argument is based on the idea that
to obtain reliability in short timescales it is necessary to average
away the large observed variability of individual ISIs by replicating
the signal through many similar neurons. In other words, the need for
rapid information transmission should strongly constrain the cortical
architecture. Our study demonstrates precisely the opposite. The fact
that the first order temporal information transmitted by a population
is simply the sum of all single-cell contributions (equation~\ref{eq:temp1stderiv}) demonstrates that it is not necessary
to transmit many copies of the same signal to ensure rapid and
reliable transmission. If each cell contributes some non-identical
information about the stimuli, this will sum up in less than one ISI,
and high population information rates can be achieved.

We finally note that an earlier work by DeWeese \cite{DeW95,DeW96} has
previously reported an elegant analytical study of the impact of
interaction between spikes of information transmission in single
cells. This study made use of a cluster-expansion formalism derived
from statistical mechanics. The results obtained by DeWeese in this
way are close to what we obtained in the single cell case. The only
difference is that in DeWeese's equations the summation over time in
the quadratic term of second derivative of the information is
restricted to pairs of time bins different from each other. However,
this work reports several advances with respect to the earlier work of
DeWeese. In fact, we computed the information carried by an ensemble
of cells, instead of by single cells only. We separated the
information into different components, each reflecting different
encoding mechanisms. We also studied the transition between spike
count and temporal encoding regimes. The power series formalism makes
the conditions under which the series converges transparent; this
issue is much less clear with a cluster expansion (see
e.g. \cite{DeW95} p. 41). Unlike in \cite{DeW96}, corrections for
finite sampling were introduced here. This last development is
essential when using the method for studying information transmission
in real neurons.

In conclusion, a full understanding of the coding properties of a
neural system cannot be achieved simply by computing the total
information about the stimuli contained in its spike trains; rather,
it is necessary to at the very least discover how this total is
comprised from the individual information-bearing parameters. The
results that we have presented here encourage us to think that this is
possible.

\section*{Appendix A}

\setcounter{equation}{0}
\def\theequation{A.\arabic{equation}}

In this Appendix we show that, under the assumptions summarized in
section \ref{sec:roval}, the only responses contributing to the
transmitted information up to order $k$ are the spike patterns with up
to $k$ spikes in total. Thus the power expansion in the short time
window length $T$ is related to the expansion in the total number of
spikes emitted by the population. We also compute the only non-zero
response probabilities up to second order, which are the only ones
contributing to the first two information derivatives. We give the
proof for finite temporal precision.

Consider the response to stimulus $s$. The probability of observing
one spike in the bin centered at $t^a_i$, averaged over all
possible patterns of spikes occurring in other bins, is $r_a(t^a_i;s)
\Delta t $. Now, note that the assumption specified by
equation~\ref{conditionalprobability} implies that the probability of
a single spike, the observation of {\em any pattern} of other spikes,
also scales proportionally to $\Delta t$:
\eq
P(t^a_i | t^b_j \cdots t^f_k ; s) \propto \Delta t
\en
Thirdly, we use the well known theorem on compound probabilities to
write down the following chain rule for the probability of observing a
pattern $t^a_i t^b_j \cdots t^f_k$ of $k$ spikes when stimulus $s$ is
presented:
\eqa
P(t^a_i t^b_j \cdots t^f_k; s) & = & P(t^a_i | t^b_j \cdots t^f_k ;s)
P(t^b_j |  \cdots t^f_k; s) \cdots P(t^f_k ; s)
\ena

This means that a probability of $k$ spikes is a product of $k$
conditional probabilities of emission of each of the spikes given the
presence of other spikes in the pattern. Since, as proven above, each
of the $k$ terms of this product is proportional to $\Delta t$,
probabilities of $k$-plets of spikes scale as $(\Delta t)^k$. From
this scaling property, from the definition of information, equation
\ref{eq:tempmutinfo}, and from the logarithm series expansion, equation
\ref{eq:logexpansion}, it follows that, for the computation of the
first $k$ information derivatives, only response probabilities with up
to $k$ spikes are needed, and they can be truncated at the $k$-th
order in $\Delta t$.  Responses with more than $k$ spikes do not
contribute to the first $k$ information derivatives.

Finally we compute the response probabilities of up to two spikes,
truncated at the second order in $\Delta t$. The probability of
observing just two spikes at $t_1^a, t_2^b$ is, up to $O(\Delta t^2)$,
given by the product of $P(t_1^a|t_2^b ;s)$ (from
equation~\ref{conditionalprobability}) and $r_b(t_2^b ;s) \Delta t$
(the latter factor quantifying at first order the probability of one
spike in $t_2^b$):
\eqa
P(t_1^a t_2^b ;s) &=& {1\over 2} \overline{r}_{a}(t_1^a
; s) \overline{r}_{b}(t_2^b ; s) \left[1 +  
\gamma_{ab}(t_1^a , t_2^b ; s)\right] (\Delta t)^2
+ O(\Delta t^3)
\label{twospikeprob}
\ena
The probability of two spikes is divided by two to prevent over-counting
due to equivalent permutations, rather than restrict the sum over
events to non-equivalent permutations, as was done in
\citeasnoun{Pan+99cor}. The probability $P(t_1^a;s)$ of observing
just one spike at $t_1^a$ is given, up to order $\Delta t^2$, by the
product of $r_b(t_2^b ;s) \Delta t$ and the probability of not
observing any other spike than that at $t_1^a$:
\eqa
P(t_1^a; s) & = & r_a(t_1^a ;s) \Delta t 
P({\rm ~no ~spikes ~but} ~ t_1^a ;s) + O(\Delta t^3) \nonumber
\\ & = & r_a(t_1^a ;s) \Delta t \left( 1 - \sum_{b=1}^C \sum_{t_2^b} 
P(t_2^b|t_1^a;s) - \sum_{b,c}\sum_{t_2^b t_3^c} P(t_2^b t_3^c | t_1^a)
- \cdots \right) 
\nonumber \\ 
& =& r_a(t_1^a ;s) \Delta t \left( 1 - 
\Delta t  
\sum_{b=1}^C \sum_{t_2^b}
\overline{r}_{b}(t_2^b ; s)
\left[1 + \gamma_{ab}(t_1^a , t_2^b ; s)\right] \right)\nonumber\\
& + & O(\Delta t^3)
\label{onespikeprob}
\ena 
The probability of no cells at all firing, up to $\Delta t^2$, can be
obtained by subtracting from 1 all the response probabilities which
are non-zero at second order.

The probability expansion in the infinite temporal precision limit can
be obtained along the same lines, by replacing probabilities with
probability densities, and using equation~\ref{eq:discrtocont}. 

\section*{Appendix B}

\setcounter{equation}{0}
\def\theequation{B.\arabic{equation}}

When considering neurophysiological data, the information quantities
have to be evaluated from response probabilities obtained from a
limited number of trials per stimulus, $N_s$. This induces an upward
systematic error, or bias, in the information measures. The systematic
error has to be evaluated and subtracted in order to obtain unbiased
information measures. In this appendix we briefly report how the bias
is computed and corrected for in the different information estimation
methods.

The bias correction is a simple formula, which is slightly different
when considering the information is calculated directly from the
Shannon formula (equations~\ref{eq:tempmutinfo} and
\ref{eq:countmutinfo}) by ``brute force'' estimation of response
frequencies, or when considering the information computed by the
series expansion\footnote{\normalsize Note that the series expansion correction
reported below is the leading bias term in the short time limit
only. Note also that the series expansion bias correction slighlty
differes from the one for the brute force computation because in the
series expansion formalism, the response class corresponding to zero
response has always non-zero probability. This response class
contribution cancels out the ``-1'' in the bias for the brute force
infromation.} (see
\citeasnoun{Pan+96},\citeasnoun{Pan+99cor}):
\eq \delta I_{\rm brute force} = \frac{\sum_{s\in \cal
S}  (R_s-1) -(R-1)}{2 N \ln 2} ;
\qquad 
\delta I_{\rm series} = \frac{\sum_{s\in \cal
S}  (R_s) -R}{2 N \ln 2}
\label{eq:bias}
\en 
where $N$ is the total number of trials (across all stimuli), and $R$
is the number of relevant response classes across all stimuli
(i.e. the number of different responses with non-zero probability of
being observed). $R_s$ is the number of relevant response classes to
stimulus s \cite{Pan+96}. It is intuitive that the number of trials
per stimulus $N_s$ should be big when compared to the number of
response classes $R$ in order to get good enough sampling of
conditional responses, and therefore reliable bias correction. When a
Bayes procedure (that takes into account that some response classes
may not be observed just because of local undersampling) is used to
estimate the number of relevant bins, reliable bias corrections can be
obtained when $N_s$ is equal or higher than the number of response
classes $R$. This gives a useful rule of thumb for evaluating the
range in which the bias corrections still work.

For the brute force full temporal information
(equation~\ref{eq:tempmutinfo}), the response space is that of all the
possible temporal spike patterns, and the number of possible responses
is $2^{C d}$, $d = T/\Delta t$ being the number of time bins in which
the window is digitised.  The situation is very different when the
information is computed by the series expansion approach.  For the
first order spike times information, $R$ is the number of non-zero
bins of the dynamic rate function, which is at most $C d$. For the
second order spike times information, $R$ is the number of relevant
bins in the space of pairs of spike firing times, which is at most $C
d (d-1) /2 \; + C (C-1) d (d+1)/4$. Therefore the number of trials
needed to effectively correct for the bias grows only linearly when
$C$ or $d$ is increased if the first order term is considered, and
only quadratically when also the second order term is
included. Considering that the growth of the data size required for
the brute force evaluation from the Shannon formula is exponential,
there is a clear advantage in terms of data size when using the series
approach developed here. Some simulation examples of effectiveness of
bias removal procedures for the series expansion method were reported
in \citeasnoun{Sch+00tempent}.

As an example, when 50-55 trials per stimulus are available for
information analysis of single cells (as in the example reported
here), the brute force computation is unbiased only up to 4-5 time
bins, and the series expansion is unbiased up to 9-10 time bins.  The
spike count information quantities of course require much less data
than the corresponding spike times quantities. They are thus well
corrected for bias for the whole range in which the full temporal
information measures are reliable.

\section*{Acknowledgements}

We thank W. Bair, P. Dayan, J. Gigg, J.A. Movshon, R. Petersen, E.T. Rolls, A. Treves and M.P. Young for useful discussions. We also thank M. Lebedev and M. Diamond for making available to us the barrel cortex data. This
research was supported by the Wellcome Trust Programme Grant
055075/Z/98 (S.P.) and by the Howard Hughes Medical Institute (S.R.S.)

\vspace{1cm}
\hrule
\normalsize
\def\refname{References}
\bibliography{tcode}

\newpage

\centerline{\bf Figure Captions}

\noindent
{\bf Figure 1:} The accuracy of the first and second order approximations to the full
information in the spike times of a single cell with Poisson responses
to two stimuli. The cell responded to stimulus 1 with a constant (in
time) rate of 30 spikes/sec., and to stimulus 2 with a spike rate
oscillating sinusoidally around 30 spikes/sec. with period 2 ms and
amplitude 30 spikes/sec.

\bigskip\noindent
{\bf Figure 2:} The scaling relationship between the average conditional firing rate (computed as the average probability of
observing a spike at one time, given that a spike has been observed at
another time, divided by the bin width $\Delta t$) and the bin width of observation. The relationship is shown for two cells from the rat barrel cortex. 

\bigskip\noindent
{\bf Figure 3:}
Comparison of the full temporal and spike count
information. (A) The modulation of the firing rates of a single
non-homogeneous Poisson neuron by two stimuli. (B) The evolution
of the full temporal and the spike count information as the time
window is increased in width from the first instance of stimulus
related responses

\bigskip\noindent
{\bf Figure 4:}
The effect of the frequency of stimulus modulation of the
rates. {\bf a} Full temporal information. {\bf b} Purely temporal
information only. The main requirement for temporal contribution to
the information can be seen easily in the comparison of these two
figures -- fast fluctuation of the firing rates in comparison to the
timescale of observation.

\bigskip\noindent
{\bf Figure 5:}
A situation in which stimulus dependent correlation in spike
timing between cells can make a sizeable (but nevertheless second
order) contribution to the total information, via component {\em 2c}
of the temporal information. (a) Correlation function. (b) Resulting
components of the full temporal information. Component {\em 2c}
dominates. Note that the corresponding spike count component was more than ten times smaller over the whole time range.

\bigskip\noindent
{\bf Figure 6:}
The effect of precision of spike timing on the information
available from 32 ms of the response of a single simulated neuron. We
used an absolute refractory period of 4 ms in this simulation (in
order to have at most one spike per time bin for all the precisions
`used). The data obtained with the continuous-time limit (obtained by
numerical integration) are included as the y-intercept information
values.

\bigskip\noindent
{\bf Figure 7:} The effect of the relative frequencies and phase of signal and noise
temporal variations in achieving synergistic coding. (A) With signal
and noise correlation oscillating with identical timescales. The line
marked with the star symbol indicates synergistic coding in this case,
as it falls above the sum of single cells' information. (B) Noise
correlation oscillating ten times faster than signal correlation.

\bigskip\noindent
{\bf Figure 8:}
Synergy between spikes, measured by the fraction of the
temporal information contained in the second order terms. (A)
Comparison of a Poisson process with processes augmented with absolute
refractory periods. In one case a relative refractory period with
exponential autocorrelation is also added (circles). (B) The
interaction of a 4 ms refractory period with stimulus-induced
modulation frequency, for three different modulation frequencies.

\bigskip\noindent
{\bf Figure 9:} {\bf (a)}: The time course of the information in the spike
counts for a neuron located in the D2 barrel column of the rat
cerebral cortex. The information obtained with the brute force method
(equation~\ref{eq:countmutinfo}, indicated with $\circ$) is compared to
that obtained with the series expansion at second order ($+$).  The
three components of the series expansion information are also
presented (rate ($\diamond$); stimulus independent correlational
(box); stimulus dependent correlational ($\star$)). {\bf (b)}: The
time course of information carried by spike times of the same neuron,
computed with a 10 ms precision. Notation as in (a). The `rate'
component in this case of course refers to the dynamic firing rate
across the time window, rather than anything to do with a spike count
code.

\bigskip\noindent
{\bf Figure 10:}
The spike count and spike times information time course for a
second neuron located in the C2 barrel column. Notations as in
Figure~\ref{fig:exam1}.

\newpage

\begin{figure}
\epsfysize=9.0truecm 
\epsfxsize=10.0truecm 
\centerline{\epsffile{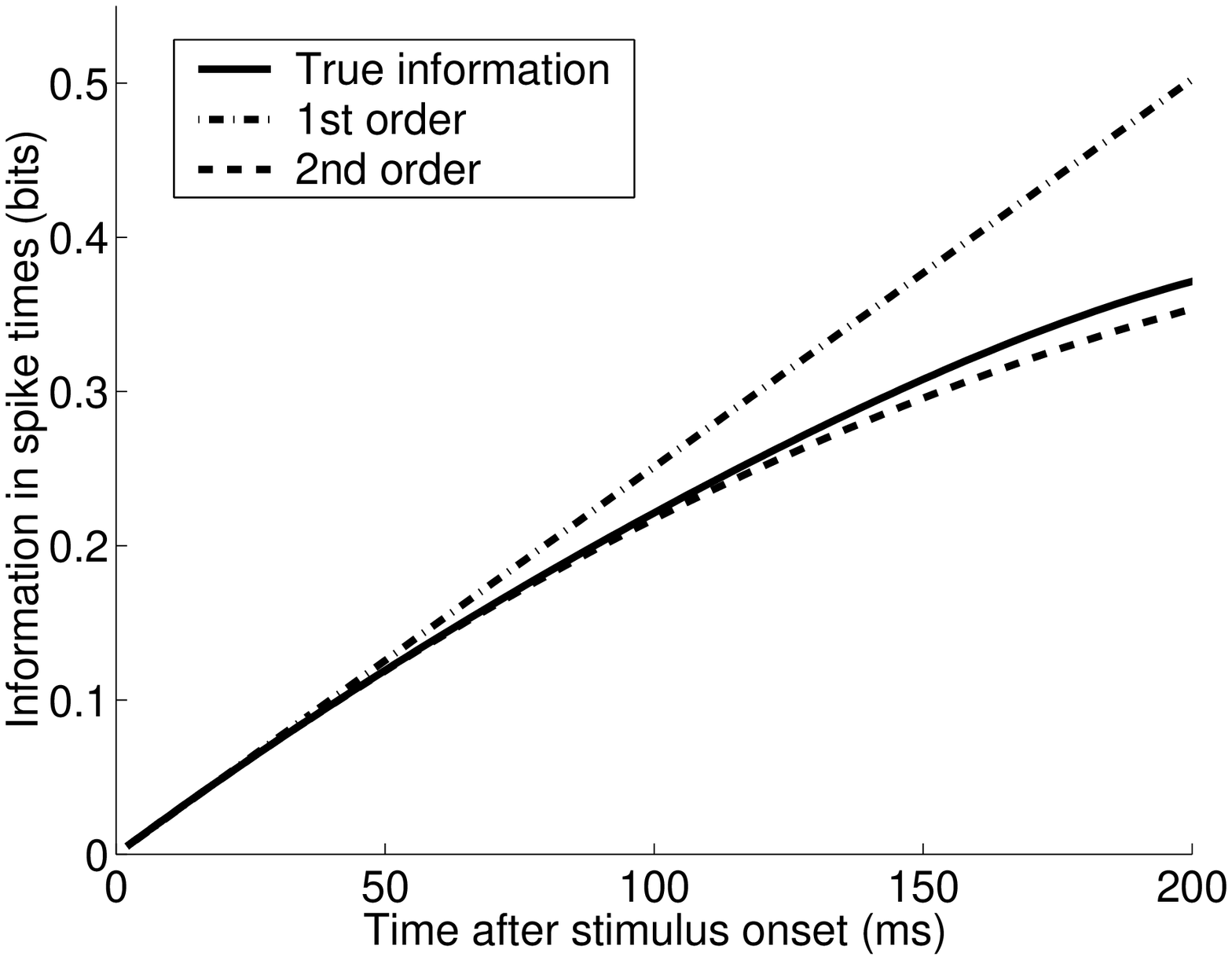}}
\vskip 3.0truecm
\caption{Panzeri, MS 2096}
\label{fig:roval1cell}
\end{figure}

\begin{figure}
\begin{center}
\leavevmode 
\epsfysize=9.0truecm 
\epsfxsize=10.0truecm 
\epsffile{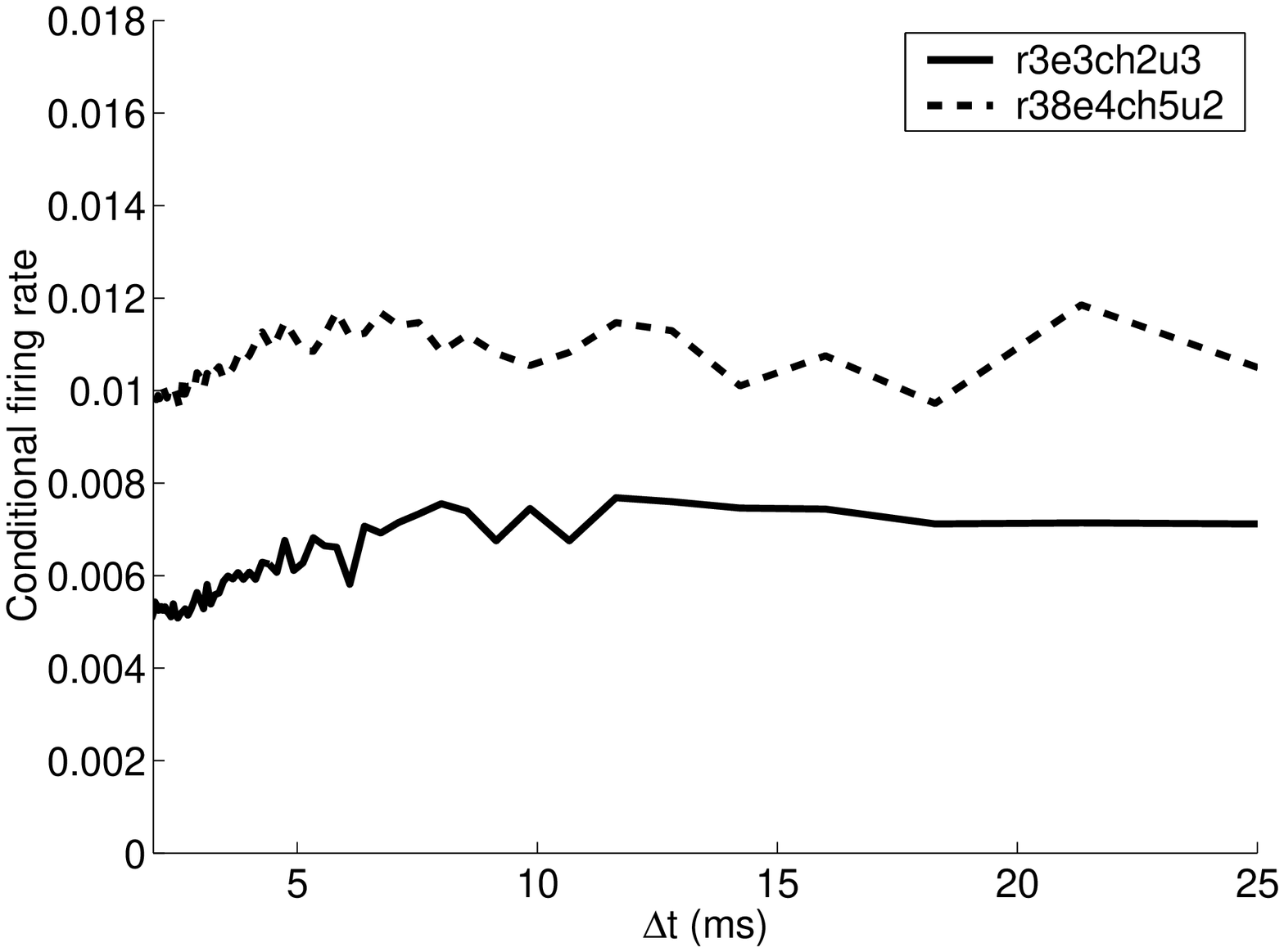}
\end{center}
\vskip 3.0truecm 
\caption{Panzeri, MS 2096}
\label{fig:scaling}
\end{figure}

\begin{figure}
\begin{center}
\leavevmode 
\epsfysize=9truecm 
\epsfxsize=10truecm 
\epsffile{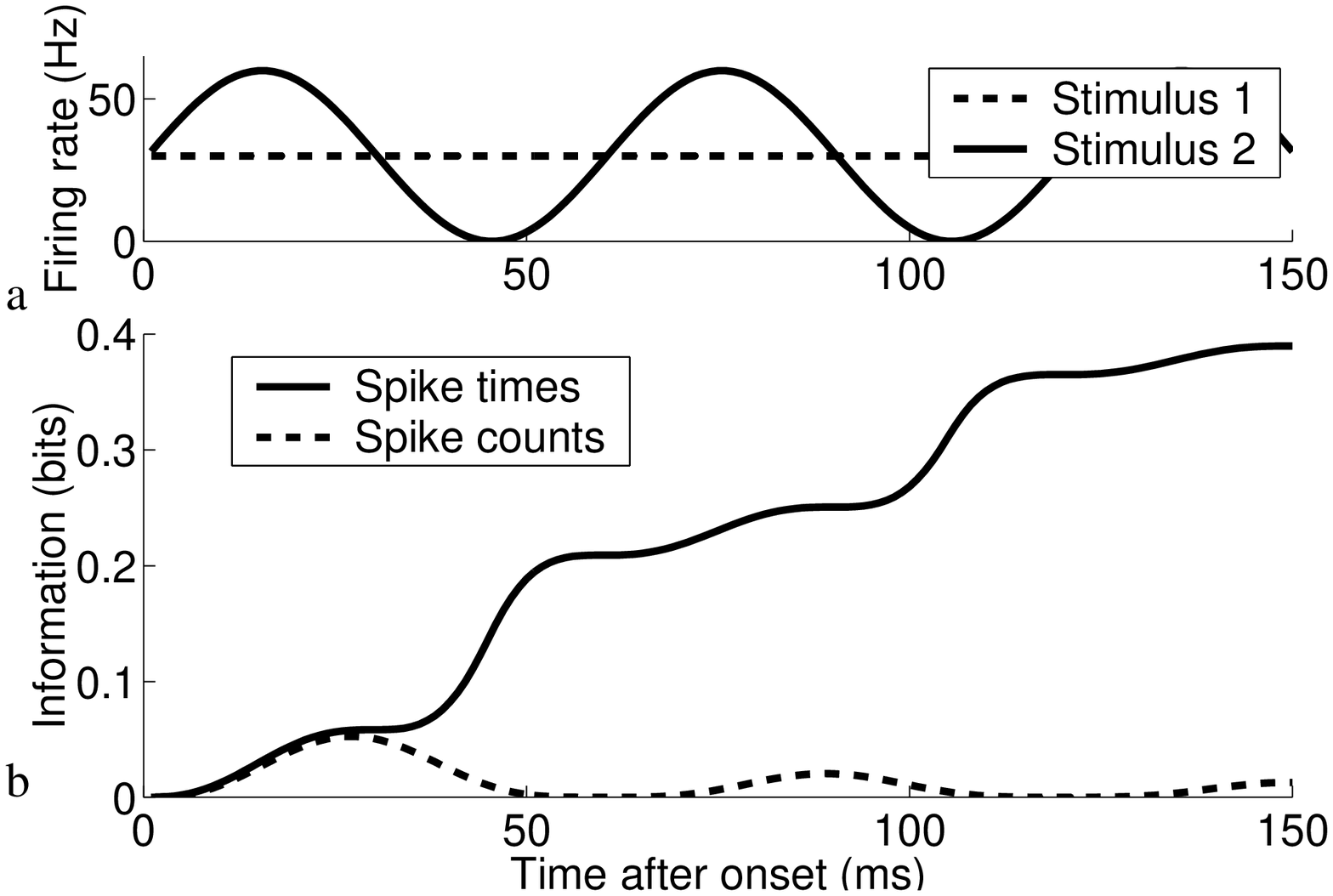}
\end{center}
\vskip 3.0truecm 
\caption{ Panzeri, MS 2096}
\label{fig:cf}
\end{figure}

\begin{figure}
\begin{center}
\leavevmode 
\epsfysize=18.0truecm 
\epsfxsize=10.0truecm 
\epsffile{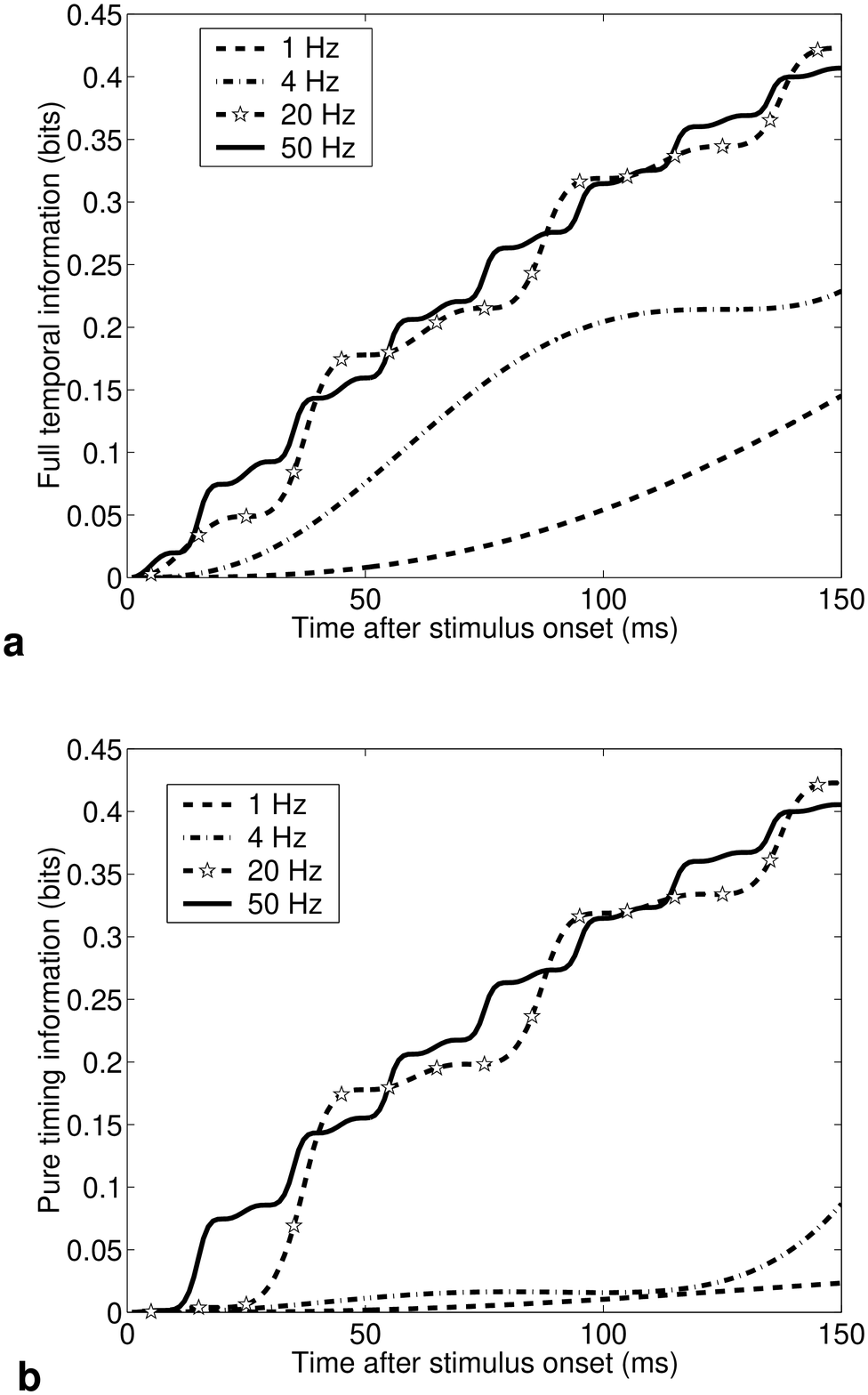}
\end{center}
\caption{ Panzeri, MS 2096}
\label{fig:freq}
\end{figure}

\begin{figure}
\begin{center}
\leavevmode 
\epsfysize=9truecm 
\epsfxsize=10truecm 
\epsffile{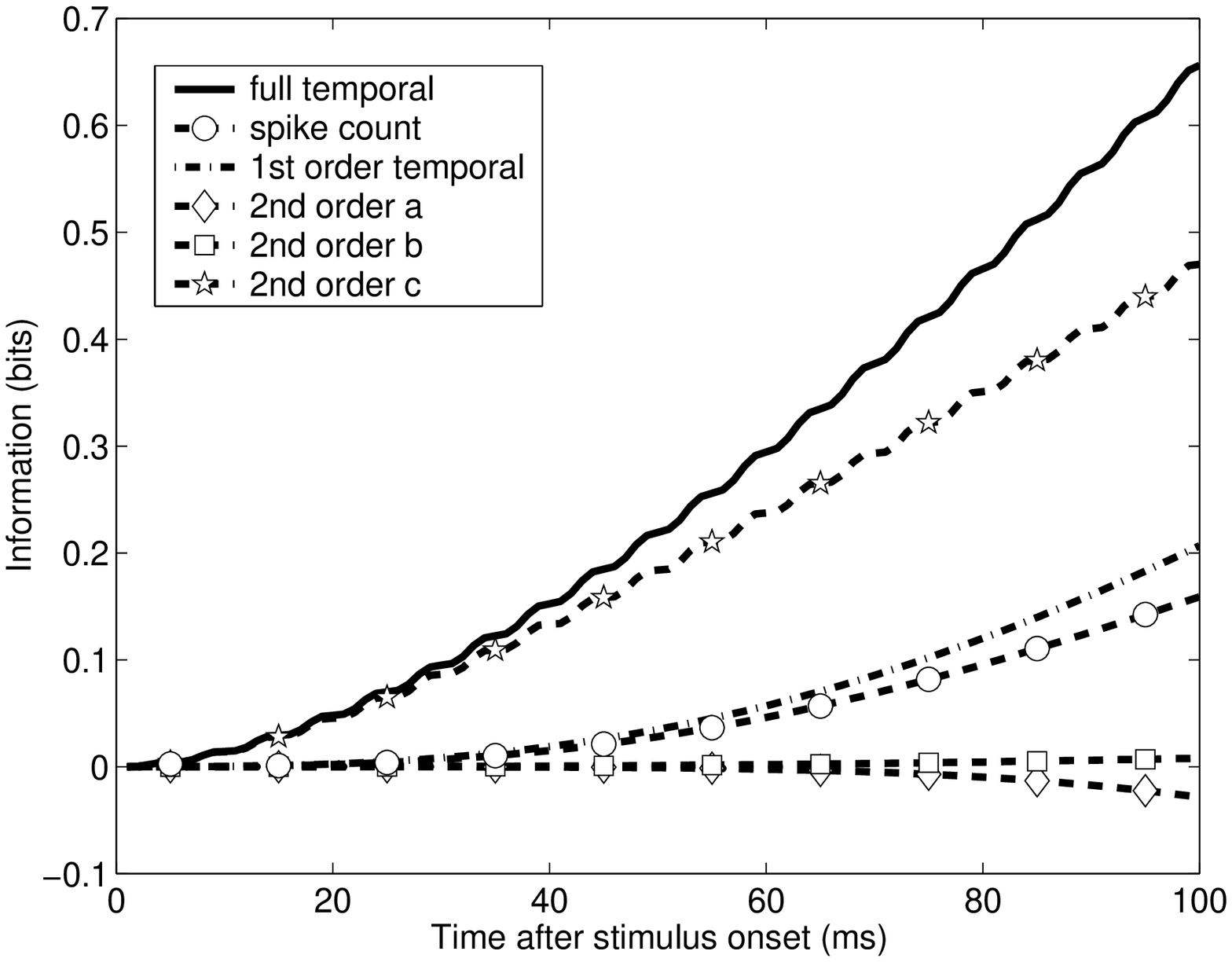}
\end{center}
\vskip 3.0 truecm
\caption{ Panzeri, MS 2096}
\label{fig:fastcor}
\end{figure}

\begin{figure}
\epsfysize=9.0truecm 
\epsfxsize=10.0truecm 
\centerline{\epsffile{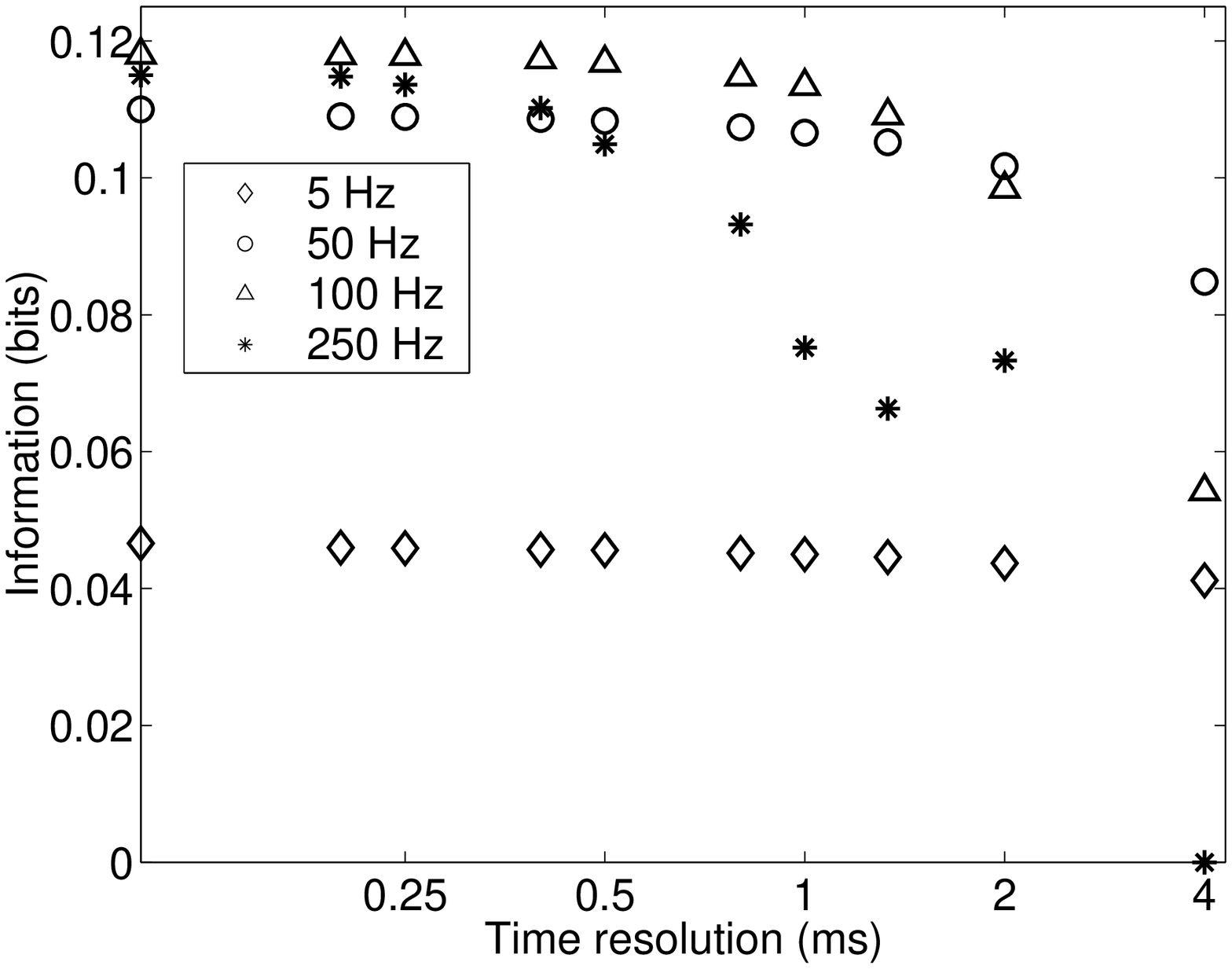}}
\vskip 3.0 truecm 
\caption{ Panzeri, MS 2096}
\label{fig:precis}
\end{figure}

\begin{figure}
\begin{center}
\leavevmode
\epsfysize=18.0truecm 
\epsfxsize=10.0truecm 
\epsffile{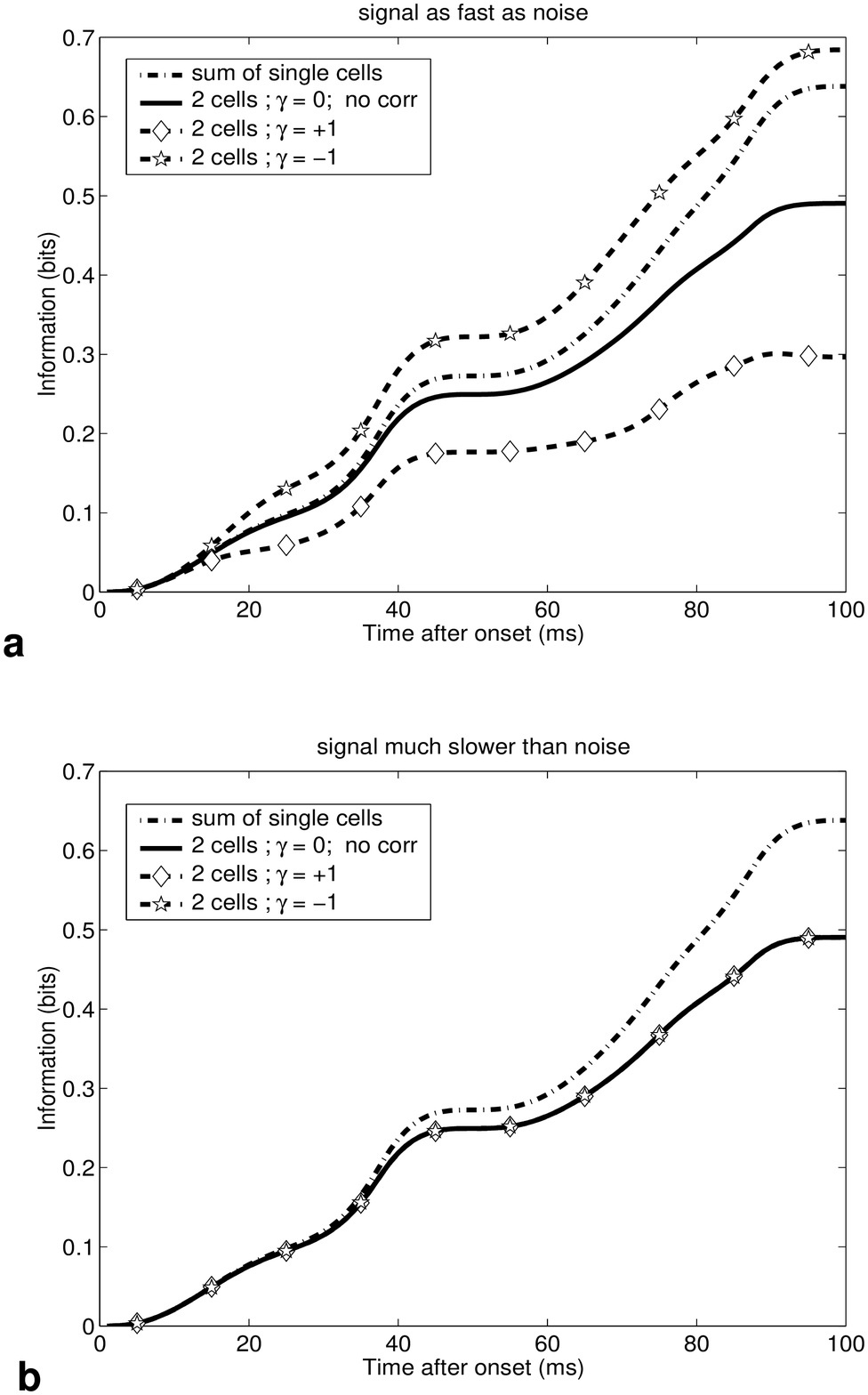}
\end{center}
\caption{ Panzeri, MS 2096}
\label{fig:popsyn}
\end{figure}

\begin{figure}
\begin{center}
\leavevmode
\epsfysize=18.0truecm 
\epsfxsize=10.0truecm 
\epsffile{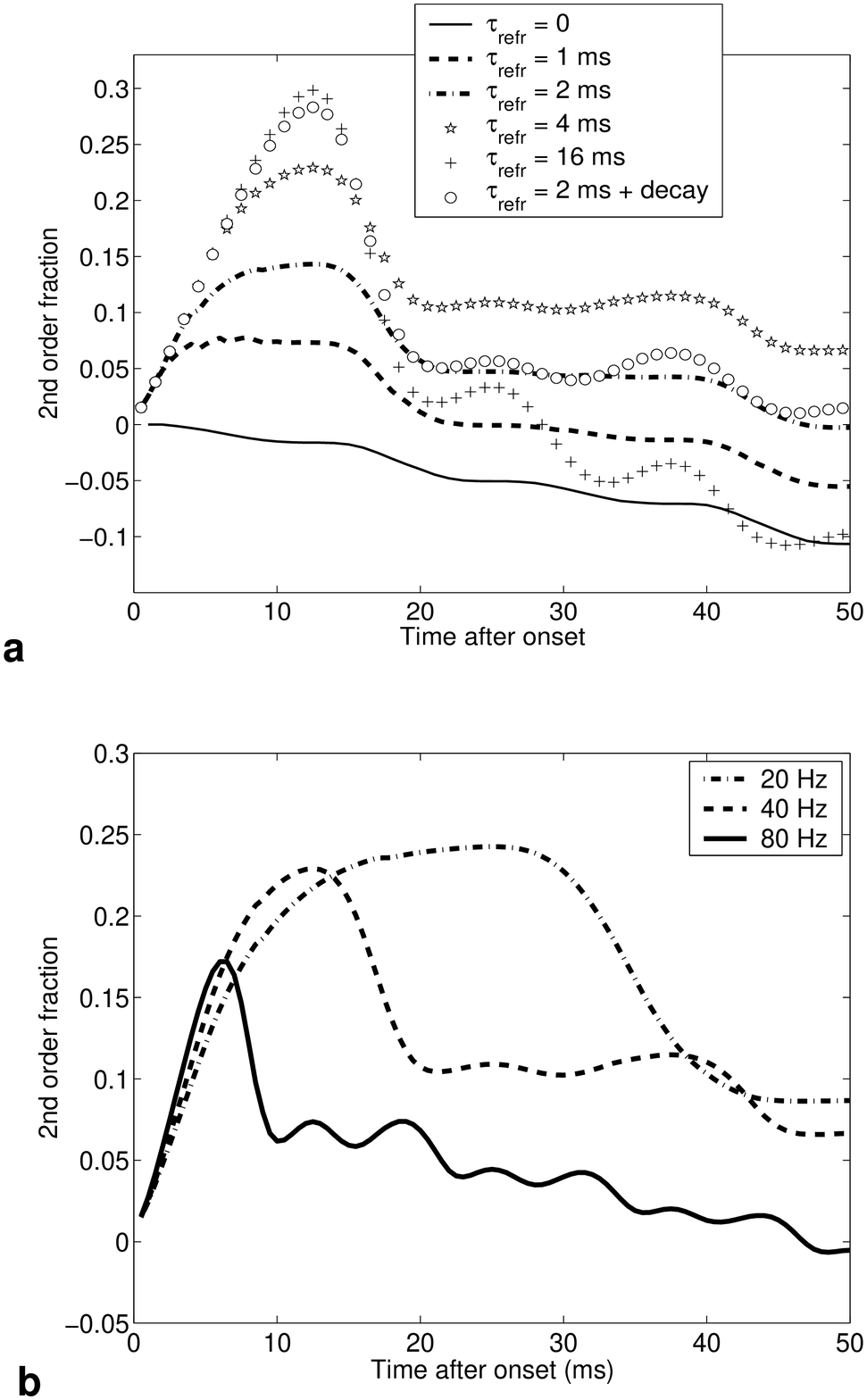}
\end{center}
\caption{ Panzeri, MS 2096}
\label{fig:spksyn}
\end{figure}

\begin{figure}
\begin{center}
\leavevmode 
\epsfysize=18.0truecm 
\epsfxsize=9.0truecm 
\epsffile{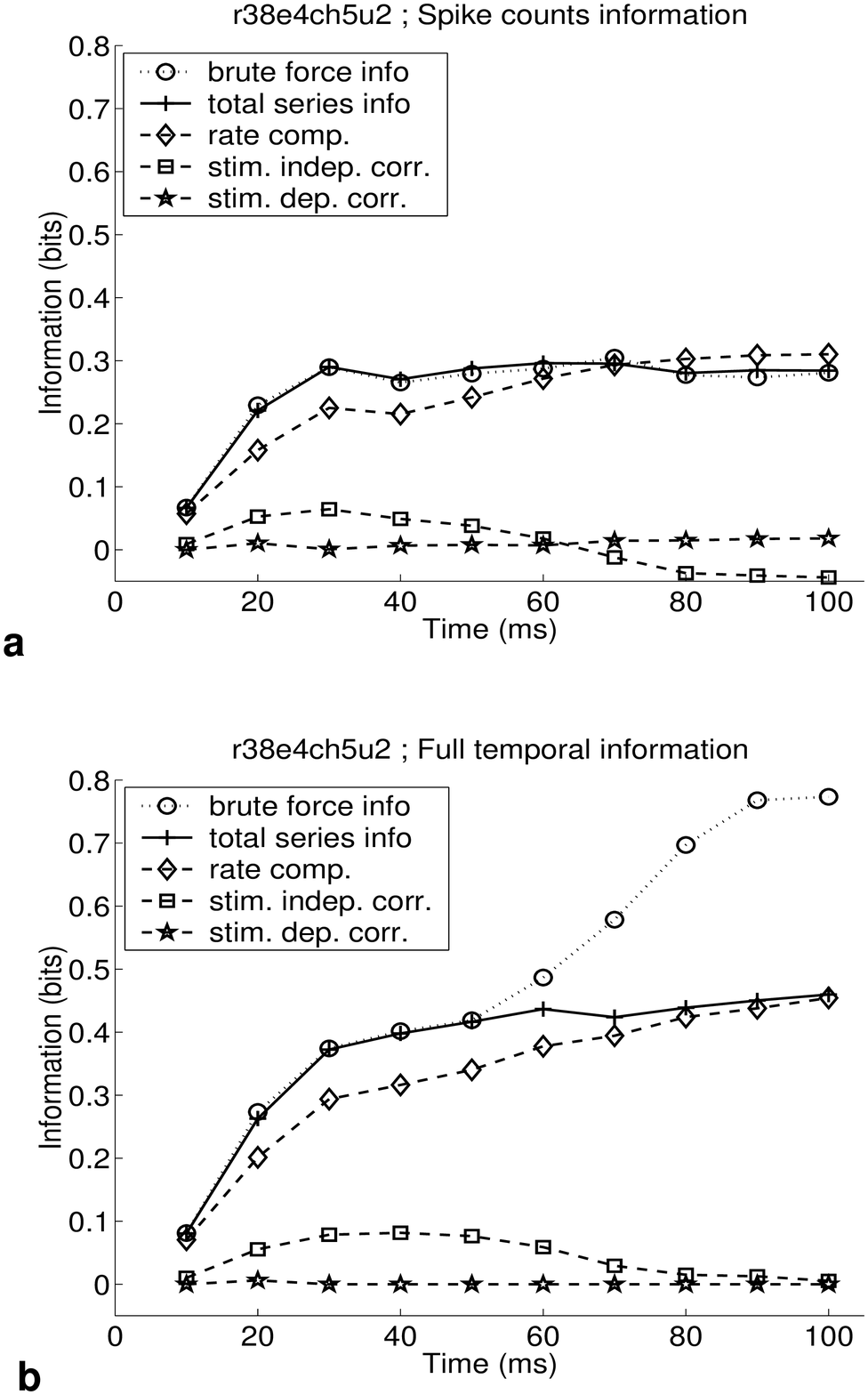}
\end{center}
\caption{ Panzeri, MS 2096}
\label{fig:exam1}
\end{figure}

\begin{figure}
\begin{center}
\leavevmode 
\epsfysize=18.0truecm 
\epsfxsize=9.0truecm 
\epsffile{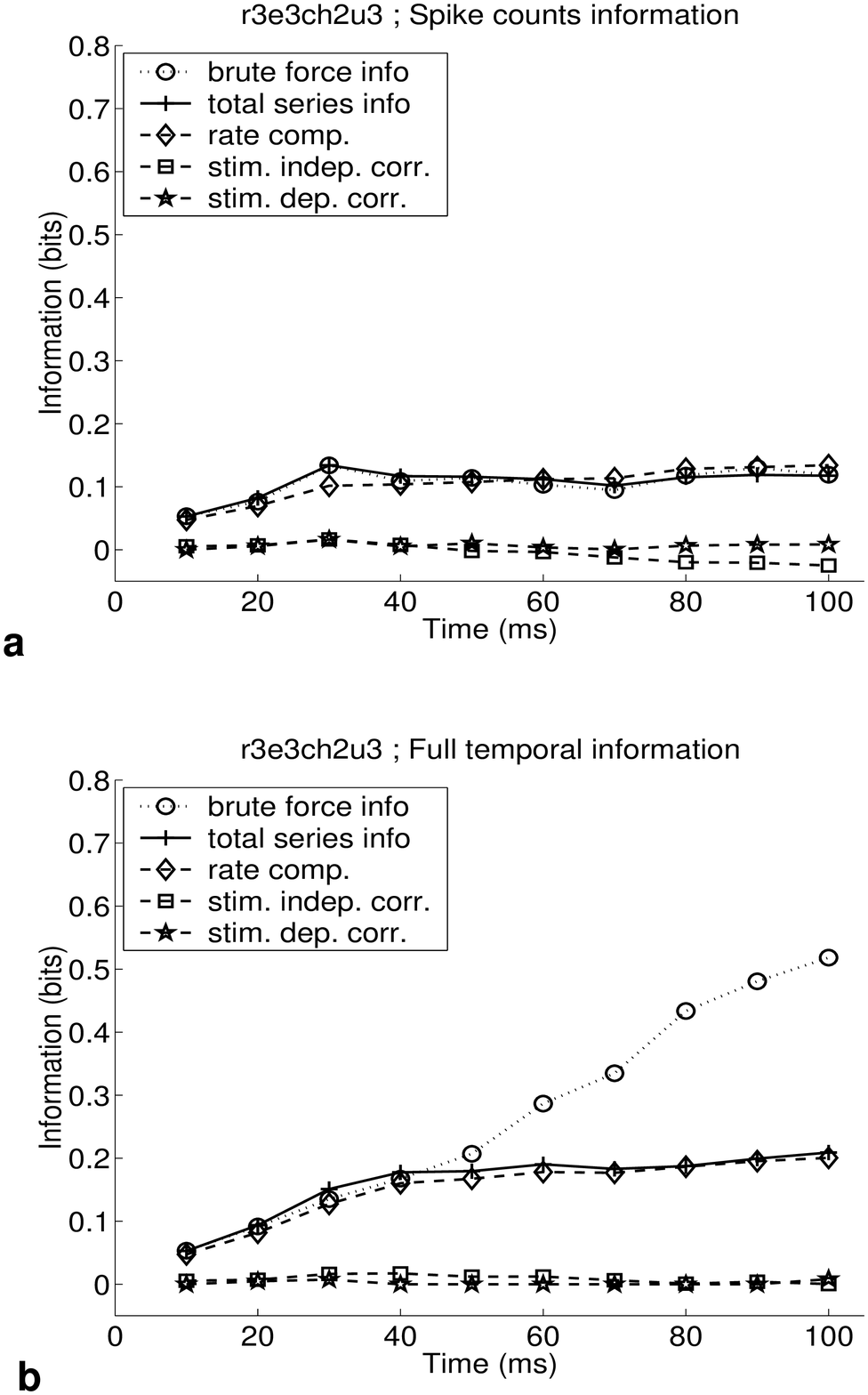}
\end{center}
\caption{ Panzeri, MS 2096 }
\label{fig:exam2}
\end{figure}

\end{document}